\documentclass[twocolumn]{aastex62}
\usepackage{amsmath,amssymb}
\usepackage{graphicx}
\usepackage{ascmac}
\usepackage{mathtools}
\usepackage{comment}
\usepackage{bm}
\usepackage{mathrsfs}
\newcommand{\p}{\partial}
\newcommand{\f}[2]{\frac{#1}{#2}}
\newcommand{\mr}[1]{\mathrm{#1}}
\newcommand{\h}{\hspace{1mm}}
\newcommand{\bra}[1]{\left(#1\right)}
\newcommand{\ctext}[1]{\raise0.2ex\hbox{\textcircled{\scriptsize{#1}}}}
\newcommand{\chH}{\mathrm{H}}
\newcommand{\chHp}{\mathrm{H}^{+}}
\newcommand{\chHm}{\mathrm{H}^{-}}
\newcommand{\chHt}{\mathrm{H}_{2}}
\newcommand{\che}{\mathrm{e}^{-}}
\DeclareRobustCommand{\plural}{\eqref{eq:Mdot_ratio_2}}

\submitjournal{ApJ}
\shorttitle{Massive disks around primordial protostars}
\shortauthors{Kimura et al.}

\begin{document}
\title{Growth of Massive Disk and Early Disk Fragmentation in the Primordial Star Formation}
\correspondingauthor{Kazutaka Kimura}
\email{kazutaka.kimura@yukawa.kyoto-u.ac.jp}
\author[0000-0001-8382-3966]{Kazutaka Kimura}
\affil{Center for Gravitational Physics, Yukawa Institute for Theoretical Physics, Kyoto University, Kyoto 606-8502, Japan}
\author[0000-0003-3127-5982]{Takashi Hosokawa}
\affiliation{Department of Physics, Graduate School of Science, Kyoto University, Sakyo, Kyoto 606-8502, Japan}
\author[0000-0001-7842-5488]{Kazuyuki Sugimura}
\affiliation{Department of Astronomy, University of Maryland, College Park, MD 20740, USA}
\begin{abstract}
Recent high-resolution simulations demonstrate that disks around primordial protostars easily fragment in the accretion phase before the protostars accrete less than a solar mass. To understand why the gravitational instability generally causes the fragmentation so early, we develop a one-dimensional (1D) non-steady model of the circumstellar disk that takes the mass supply from an accretion envelope into account. We also compare the model results to a three-dimensional (3D) numerical simulation performed with a code employing the adaptive mesh refinement. Our model shows that the self-gravitating disk, through which the Toomre $Q$ parameter is nearly constant at $Q \sim 1$, gradually spreads as the disk is fed by the gas infalling from the envelope. We further find that the accretion rate onto the star is an order of magnitude smaller than the mass supply rate onto the disk. This discrepancy makes the disk more massive than the protostar in an early evolutionary stage. Most of the infalling gas is used to extend the outer part of the self-gravitating disk rather than transferred inward toward the star through the disk.
We find that similar evolution also occurs in the 3D simulation, where the disk becomes three times more massive than the star before the first fragmentation occurs. Our 1D disk model well explains the evolution of the disk-to-star mass ratio observed in the simulation. We argue that the formation of such a massive disk leads to the early disk fragmentation.
\end{abstract}
\section{Introduction} \label{sec:intro}
Primordial stars, also known as Population III stars, are formed in the pristine universe after the Big Bang \citep[e.g.,][]{Greif_2015,Haemmerle_et_al_2020}.
Their radiation heats and reionizes the surrounding medium, and their supernova explosion enriches primordial gas with heavy elements.
Elucidating primordial star formation is thus necessary to understand their roles in the subsequent structure formation.
\par
The primordial star formation begins with the gravitational collapse of a massive cloud with $\sim 1000$ $M_\odot$ \citep{Abel_et_al_2002,Bromm_et_al_2002}.
A tiny protostar appears after the collapse, and it rapidly grows accreting the gas through a circumstellar disk at typical rates $\dot{M}_* \sim 10^{-3}~\text{--}~10^{-2}~M_\odot {\rm yr}^{-1}$ \citep{Omukai_and_Nishi_1998}.
Many simulations have shown that circumstellar disks fragment and then binary and multiple systems emerge in the protostellar accretion stage  \citep{Saigo_et_al_2004,Machida_et_al_2008,Turk_et_al_2009,Stacy_et_al_2010,Greif_et_al_2011,Susa_2019}.
In particular, high-resolution simulations show disks fragment when the protostellar mass is less than a solar mass \citep{Clark_et_al_2011,Greif_et_al_2012}.
The stellar mass growth still continues until the radiative feedback halts the accretion \citep{Omukai_and_Palla_2003,McKee_and_Tan_2008,Hosokawa_et_al_2011}, when the stellar mass is several to a thousand solar masses \citep[e.g.,][]{Hirano_et_al_2014,Susa_et_al_2014,Hirano_et_al_2015,Stacy_et_al_2016, Sugimura_et_al_2020}.
\par
There is another pathway for primordial star formation, the so-called direct collapse scenario \citep[][]{Bromm_and_Loeb_2003}.
The direct collapse postulates that some physical processes provide a massive primordial cloud that collapses via H atomic cooling \citep{Omukai_2001}. A protostar then grows in mass at very high accretion rates of $\dot{M}_* \sim 0.1~\text{--}~1~M_\odot {\rm yr}^{-1}$. If such very rapid accretion continues for $\sim$ 1~Myr,  supermassive stars with $10^{5}~\text{--}~10^{6}~M_\odot$ form and they may provide massive seeds of supermassive black holes after they die \citep[see][for reviews]{Volonteri_2010, Inayoshi_et_al_2020}.
However, previous simulations also show that the circumstellar disks form \citep[e.g.,][]{Latif_et_al_2013} and fragment also for this case  \citep{Becerra_et_al_2015,Sakurai_et_al_2016,Chon_et_al_2018,Latif_et_al_2020,Matsukoba_et_al_2020}. If too many fragments appear, they might evolve into a cluster of lower-mass stars instead of a single supermassive star \citep[e.g.,][]{Regan_et_al_2018}.
Moreover, disk fragmentation leads to the substantial variability of the accretion rate as fragments migrate toward the central star.
Such variability affects the protostellar evolution and may  strengthen the radiative feedback to prevent the supermassive star formation
\citep[][]{Sakurai_et_al_2015,Inayoshi_and_Haiman_2014}.
\par
Fragments in disks follow different evolution paths.
Whereas some rapidly migrate inward and merge with the central star, others migrate outward and survive in the disk for a long time \citep[e.g.,][]{Hirano_and_Bromm2017,Chon_and_Hosokawa_2019}.
If binaries with small separations finally survive, they may evolve into black hole binaries that merge within the Hubble time emitting gravitational waves \citep{Kinugawa_et_al_2014,Hartwig_et_al_2016}.
Moreover, if disk fragmentation provides star-cluster environments, the N-body interaction operates to shrink the binary orbits \citep[e.g.,][]{Hong_et_al_2018,Kumamoto_et_al_2020,Tanikawa_et_al_2020,Liu_et_al_2020}. The N-body interaction also causes the ejection of stars from the cluster. In particular, if a star less massive than $0.8~M_\odot$ is ejected, it remains on the main sequence even in the present-day universe \citep{Ishiyama_et_al_2016}.
\par
For studying the disk fragmentation, multi-dimensional simulations are powerful but computationally expensive. Alternatively, one-dimensional (1D) semi-analytical models allow high-resolution and long-term calculations for various parameters with a small computational cost.
Many authors have taken this approach to investigate the disk stability in both the ordinary primordial star formation \citep{Tan_and_McKee_2004, Tanaka_and_Omukai_2014,Takahashi_and_Omukai_2017,Liao_and_Turk_2019} and the direct collapse cases \citep{Inayoshi_and_Haiman_2014,Latif_and_Schleicher_2015,Matsukoba_et_al_2019}.
A limitation of these models is assuming the steady states with constant accretion rates throughout the disks which are equal to the mass supply rate from the envelope.
Furthermore, comparisons between 1D models and three-dimensional (3D) simulations have also been limited.
It is thus still uncertain how well the 1D models approximate the evolution observed in 3D simulations.
For instance, previous 3D simulations generally show that the disk fragmentation occurs in an early accretion stage, regardless of the normal primordial and direct collapse cases.
No previous 1D models address why the gravitational instability generally causes the disk fragmentation so early in 3D.
\par
It is broadly accepted that, in the present-day star formation, a protostar accretes the gas through a circumstellar disk \citep[e.g.,][]{Hartmann1998, Stahler_and_Palla_2008,Bate_2010,Inutsuka_et_al_2010,Machida_and_Matsumoto_2011}.
Moreover, recent observations are revealing direct images of the disk fragmentation occurring in some systems \citep{Tobin_et_al_2016,Ilee_et_al_2018,Johnston_et_al_2020}. Along with the multi-dimensional numerical simulations, semi-analytical modeling of the formation and growth of the protostellar disk has been developed \citep[e.g.,][]{Cassen_and_Moosman_1981}.
For instance, \citet{Takahashi_et_al_2013} develop a non-steady disk model incorporating the mass supply from the surrounding envelope onto the disk. They confirm the validity of the model by comparing it with the 3D simulation starting from the same initial condition \citep{Machida_et_al_2010}.
\par
In this work, we develop a 1D non-steady semi-analytical model of the accretion disks growing around primordial protostars. We implement cooling processes and chemical reactions essential for the primordial star formation into the model of \citet{Takahashi_et_al_2013}.
We focus on the early accretion stages where the stellar mass is less than $10~M_\odot$ and study how the disk evolves under the rapid mass supply from the surrounding envelope.
To verify our model, we newly perform a high-resolution 3D simulation utilizing a code employing the adaptive mesh refinement \citep{Sugimura_et_al_2020}, where the disk fragmentation occurs before the protostar accretes the gas of $1~M_\odot$.
We compare the evolution in the 1D model and 3D simulation, both of which start from the same initial condition.
\par
The rest of the paper is organized as follows. We describe our non-steady model in Section \ref{sec:model}. We then present the obtained evolution of the disk varying model parameters in Section \ref{sec:result}. Comparisons between the model and the simulation are given in Section \ref{sec:comparison}. We discuss the implication of our results in Section \ref{sec:discussion} and give summary and conclusion in Section \ref{sec:summary}.
\section{MODEL} \label{sec:model}
In this section, we describe our 1D non-steady disk model for primordial star formation. We explain how we model the mass supply onto the disk in Section \ref{sec:accretion_from_envelope}, and present the governing equations of the disk in Section \ref{sec:equation_disk}. In Section \ref{sec:num_model} we describe how to solve the equations numerically and what model parameters we calculate for.
\subsection{Mass Supply from Envelope onto Disk}
\label{sec:accretion_from_envelope}
We model the mass supply onto the disk following \citet{Takahashi_et_al_2013}, adding some modifications. They take the Bonnor-Ebert sphere as the initial condition and estimate when and where the gas infalls onto the disk by solving the equation of motion approximately.
While their method provides a reasonable approximation for the mass supply onto the disk in a late accretion stage, we find it substantially underestimates the rate during $\sim 100$ years after the birth of a protostar. We instead assume the envelope structure consistent with the early run-away collapse, which occurs in a self-similar fashion \citep{Omukai_and_Nishi_1998}.
\par
With a given equation of state $P=K\rho^\gamma$, the envelope structure at the protostellar birth is approximated as \citep{Larson_1969,Penston_1969}:
\begin{eqnarray}
\rho &=& C_1 K^{1/(2-\gamma)} r^{-2/(2-\gamma)} \h , \label{eq:Larson_rho} \\
v &=& C_2 K^{1/2(2-\gamma)} r^{(1-\gamma)/(2-\gamma)} \h , \label{eq:Larson_v}
\end{eqnarray}
where $r$ is the radius, $\rho$ is the gas mass density, $v$ is the radial velocity, $\gamma$ is the effective polytropic index, and $C_1$, and $C_2$ are constants.
We determine $K$, $\gamma$, $C_1$, and $C_2$ so that the resulting envelope structure matches the result of a 1D hydrodynamical calculation by \citet{Omukai_and_Nishi_1998}, who follow the gravitational collapse of a primordial cloud.
During the collapse, the gas temperature gradually rises due to compressional heating counteracted by the H$_2$ molecular cooling.
The resulting thermal evolution is approximated as
\begin{eqnarray}
 T \simeq 200 \bra{\f{n_{\chH}}{10^4 \mr{cm}^3}}^{0.1} \mr{K} \h , \label{eq:TE}
\end{eqnarray}
where $n_{\chH}$ is the number density of hydrogen nuclei, which is related to the mass density $\rho$ as
\begin{eqnarray}
  n_{\rm{H}} = \f{\rho}{(1+4y_{\rm{He}})m_{\rm{H}}} \h , \label{eq:nH_rho_relation}
\end{eqnarray}
where $y_{\rm{He}}=8.333\times10^{-2}$ is the number fraction of He relative to hydrogen nuclei, and $m_{\mr{H}}$ is the mass of hydrogen nucleus.
Equation \eqref{eq:TE} implies that $\gamma$ is 1.1 $(T~\propto~\rho^{\gamma-1})$.
The parameter values that approximate the numerical result are $K=4.2 \times 10^{11}$ (hereafter referred to $K_{\mr{fid}}$), $C_1 = 6.1\times10^{7}$, $C_2 = -7.0$ (in cgs), and $\gamma=1.1$.
In this fiducial case, Equations \eqref{eq:Larson_rho} and \eqref{eq:Larson_v} are rewritten as
\begin{eqnarray}
  \label{eq:env_rho}
 \rho &=& 2.6\times10^{-9} \bra{\f{K}{K_{\mr{fid}}}}^{10/9} \bra{\f{r}{1\mr{AU}}}^{-20/9} \hspace{2mm} \mr{g/cm}^{3} \h , \label{eq:env_v} \\
  v &=& -6.9\times10^{5} \bra{\f{K}{K_{\mr{fid}}}}^{5/9} \bra{\f{r}{1\mr{AU}}}^{-1/9} \hspace{2mm} \mr{cm/s} \h .
\end{eqnarray}
To investigate how disk evolution varies with different mass supply rates from the envelope, we also consider cases where $K = 0.1~K_{\rm{fid}}$ and $K = 10~K_{\mr{fid}}$ (see Section~\ref{sec:acc}).
\par
Assuming that the rotation of the cloud is weak, we estimate the elapsed time $t$ until the gas on a spherical shell with the original radius $r$ reaches the disk as
\begin{eqnarray}
t \simeq \f{r}{v(r)} \h . \label{eq:t_infall}
\end{eqnarray}
Using the relation $r = r(t)$ derived from Equation \eqref{eq:t_infall}, the total mass supply rate from the envelope to the disk is given by
\begin{eqnarray}
  \dot{M}_{\mr{e,tot}}(t) = 4 \pi \rho(r) r^2 \f{dr}{dt} \h . \label{eq:Mdot_e_tot}
\end{eqnarray}
\par
We assume that the initial rotational velocity in the envelope is $f_{\mr{Kep}}$ times the Kepler values independent of the radius:
\begin{eqnarray}
\Omega_0=f_{\mathrm{Kep}}\Omega_{\mathrm{Kep}}=f_{\mathrm{Kep}}\sqrt{\f{GM_r}{r^3}} \h ,
\label{eq:env_Omega}
\end{eqnarray}
where $\Omega_0$ is the angular velocity and $M_r$ is the enclosed mass of the envelope inside the radius $r$.
According to previous 3D simulations starting from cosmological initial condition \citep[e.g.,][]{Abel_et_al_2002, Yoshida_et_al_2006}, the typical value of $f_{\mr{Kep}}$ is $\simeq 0.5$.
To investigate how the disk structure depends on the envelope rotation, we consider the cases where $f_{\mr{Kep}}=0.3,0.5$, and $0.8$ (see Section~\ref{sec:rot}).
\par
The mass supply at the total rate $\dot{M}_{\rm{e,tot}}(t)$ is distributed into different parts of the disk, depending on the angular momentum $j$ of the infalling gas.
The angular momentum $j$ depends on the polar angle $\theta$ ($0 \le \theta \le \pi/2$) of the original position as below
\citep[see Figure 2 in][]{Takahashi_et_al_2013}:
\begin{eqnarray}
  j &=& r^2 \sin^2 \theta \h \Omega_0 \h .
\end{eqnarray}
We assume that $j$ is conserved until the gas lands on the disk surface. We consider the mass supply rate only inside a given radius $\varpi$ of the disk, $\dot{M}_{\mr{e}}(\varpi,t)$. The local rate at the radius $\varpi$ is obtained by the $\varpi$-derivative of $\dot{M}_{\mr{e}}(\varpi,t)$,
\begin{eqnarray}
  \f{\p\dot{M}_{\mr{e}}(\varpi,t)}{\p \varpi} &=& \f{\p j}{\p \varpi}\f{\p\dot{M}_{\mr{e}}(j,t)}{\p j} \nonumber \\
  &=& 2\f{\p j}{\p \varpi}\f{\p \theta}{\p j}\f{\p\dot{M}_{\mr{e}}(\theta,t)}{\p \theta} \nonumber \\
  &=& \f{\dot{M}_{\mr{e,tot}}}{2 \Omega_0 r^2}\bra{1-\f{j}{\Omega_0 r^2}}^{-1/2} \f{\p j}{\p \varpi} \h \label{eq:Mdot_e_varpi},
\end{eqnarray}
where we have substituted the relation
\begin{eqnarray}
  \f{\p\dot{M}_{\mr{e}}(\theta,t)}{\p \theta} &=& \f{\sin\theta}{2}\dot{M}_{\mr{e,tot}} \h .
\end{eqnarray}
\subsection{Disk Evolution} \label{sec:equation_disk}
Next, we describe the governing equations of the disk.
We take a cylindrical coordinate system ($\varpi,\phi,z$) hereafter, assuming the axial symmetry.
We also assume that the disk rotates at Kepler velocity, i.e., the angular velocity $\Omega$ and specific angular momentum $j$ are given by
\begin{eqnarray}
  \Omega &=& \sqrt{\f{GM_\varpi}{\varpi^3}} \h , \label{eq:disk_Omega} \\
  j &=& \varpi^2 \Omega = \sqrt{\varpi GM_\varpi} \h \label{eq:disk_j},
\end{eqnarray}
where $M_\varpi=M_*+\int_{\varpi_\mr{in}}^{\varpi} 2 \pi \varpi \Sigma d\varpi$ is the enclosed mass of the disk inside the radius $\varpi$, $M_*$ is the protostellar mass, and $\varpi_{\mr{in}}$ is the inner boundary radius of the disk.
\subsubsection{Mass and Angular Momentum Conservation}
The mass and angular momentum conservation are given as follows:
\begin{eqnarray}
  && \f{\p}{\p t} (2\pi \varpi \Sigma) - \f{\p \dot{M}_{\mr{d}}}{\p \varpi} = \f{\p \dot{M}_{\mr{e}}}{\p \varpi} \h , \label{eq:continuity} \\
  && \f{\p}{\p t} (2 \pi \varpi \Sigma j) - \f{\p}{\p \varpi} (\dot{M}_{\mr{d}}j) = \nonumber \\
  && \hspace{5mm} \f{\p}{\p \varpi} \left((2\pi \varpi\Sigma)\nu\varpi^2\f{\p \Omega}{\p \varpi}\right)
  + \f{\p \dot{M}_{\mr{e}}}{\p \varpi} j \h , \label{eq:angular}
\end{eqnarray}
where $\dot{M}_{\mr{d}} \equiv - 2\pi \varpi \Sigma v_\varpi$ is the accretion rate through the disk, $\Sigma$ is the surface density, $v_\varpi$ is the radial velocity, and $\nu$ is the viscosity coefficient.
By subtracting the product of Equation \eqref{eq:continuity} and $j$ from Equation \eqref{eq:angular}, we obtain
\begin{eqnarray}
 2\pi\varpi\Sigma\f{\p j}{\p t} - \dot{M}_{\mr{d}}\f{\p j}{\p \varpi} = \f{\p}{\p \varpi} \left((2\pi \varpi\Sigma)\nu\varpi^2\f{\p \Omega}{\p \varpi}\right) \h . \label{eq:Md_deriv}
\end{eqnarray}
Here, from Equation \eqref{eq:disk_j} $\p j/\p t$ and $\p j/\p \varpi$ are described as
\begin{eqnarray}
  \f{\p j}{\p t} &=& \f{1}{2}\sqrt{\f{\varpi G}{M_\varpi}}\f{\p M_\varpi}{\p t} = \f{1}{2}\sqrt{\f{\varpi G}{M_\varpi}} \bra{\dot{M}_{\mr{d}} + \dot{M}_{\mr{e}}} \h , \\
  \f{\p j}{\p \varpi} &=& \f{1}{2}\sqrt{\f{GM_\varpi}{\varpi}} \bra{1+\f{\varpi}{M_\varpi} \f{\p M_\varpi}{\p \varpi}} \nonumber \\
  &=& \f{1}{2}\sqrt{\f{GM_\varpi}{\varpi}} \bra{1+\f{2\pi\varpi^2\Sigma}{M_\varpi}} \h
  \label{eq:djdpi} .
\end{eqnarray}
With the above Equations \eqref{eq:Md_deriv}~--~\eqref{eq:djdpi}, we finally get
\begin{eqnarray}
  \dot{M}_{\mr{d}} &=& -2 \sqrt{\f{\varpi}{GM_\varpi}} \f{\p}{\p \varpi} \left\{(2\pi \varpi\Sigma)\nu \varpi^2\f{\p \Omega}{\p \varpi}\right\} \nonumber \\
  && + 2\pi \Sigma \f{\varpi^2}{M_\varpi} \dot{M}_{\mr{e}} \h .
  \label{eq:Mdot_d}
\end{eqnarray}
In Equation \eqref{eq:Mdot_d}, the first term on the right-hand side represents the mass flux by the viscosity. The second term represents the accretion caused by the mass supply from the envelope. Recall that, at a given radius $\varpi$, the enclosed disk mass $M_\varpi$ increases at the supply rate $\dot{M}_{\mr{e}}$. Since the centrifugal force with the fixed specific angular momentum no longer balances with the strengthened gravity, the gas moves inward even if no viscosity operates in the disk.
\par
Regarding the viscosity coefficient $\nu$, we adopt the $\alpha$ prescription with which the gravitational torque is modeled as a function of Toomre $Q$ \citep{Shakura_and_Sunyaev_1973}:
\begin{eqnarray}
  \nu &=& \alpha c_s H = \alpha \f{c_s^2}{\Omega} \h , \label{eq:nu}\\
  \alpha &=&   A \exp (-BQ^4) \label{eq:alpha} \h ,\\
  Q &=& \f{c_s \Omega}{\pi G \Sigma} \h , \label{eq:Q_def} \\
  c_s &=& \sqrt{\f{\gamma k_B T}{\mu m_\chH}} \h , \label{eq:c_s}
\end{eqnarray}
where $H \equiv c_s/\Omega$ is the scale height, $T$ is the midplane temperature, $k_B$ is the Boltzmann constant, $\mu$ is the mean molecular weight, and $A$ and $B$ are arbitrary non-dimensional constants. The functional form of Equation \eqref{eq:alpha} represents the fact that the gravitational torque operates efficiently only when $Q \lesssim 1.4$ \citep[e.g.,][]{Boley_et_al_2006}, and it has been broadly used in previous studies \citep[e.g.,][]{Zhu_et_al_2010,Takahashi_et_al_2013}. We set $(A,B)=(1,1)$ throughout this paper. We have confirmed that the disk evolution is almost the same even if we vary $A$ and $B$ within the range of $0.1~\text{--}~10$.
\subsubsection{Vertical Structure}
We assume the hydrostatic equilibrium in the vertical direction $z$,
\begin{eqnarray}
  \f{\p p}{\p z} \simeq - \rho\f{\p \Phi_g}{\p z} \simeq - \rho \f{GM_\varpi}{\varpi^2}\f{z}{\varpi} \h ,
\end{eqnarray}
where $\Phi_{g}$ is the gravitational potential and we also assume that the disk is geometrically thin, $z \ll \varpi$. We approximate the density and pressure stratification with the isothermal equation of state $p=c_s\rho$ as
\begin{eqnarray}
  \rho(\varpi,z) &=& \rho_0 \exp \left(-\f{z^2}{2H^2}\right) \h , \label{eq:rho_z} \\
  p(\varpi,z) &=& p_0 \exp \left(-\f{z^2}{2H^2}\right) \h ,
\end{eqnarray}
where quantities with subscript 0 denote those at the disk midplane. Integrating Equation \eqref{eq:rho_z} in the z direction provides
\begin{eqnarray}
  \rho_0 &=& \f{\Sigma}{H\sqrt{2\pi}} = \f{\Sigma\Omega}{c_s \sqrt{2\pi}} \h . \label{eq:iso-Sigmarho}
\end{eqnarray}
We use this to evaluate the mass density in the disk midplane, $\rho_0$.
\subsubsection{Energy Conservation}
To calculate the temperature $T$, which is necessary to evaluate the viscosity $\nu$ (see Equations~\ref{eq:nu}~--~\ref{eq:c_s}), we solve the energy conservation equation taking account of the mass supply from the envelope, thermal processes, and chemical reactions of the primordial gas:
\begin{eqnarray}
  \f{\p}{\p t} (\Sigma e_{\mr{th}})
  &=& \f{1}{2\pi\varpi} \f{\p}{\p \varpi} \left[\dot{M}_{\mr{d}} \bra{\rho e_\mr{th}+\f{p}{\rho}}\right] \nonumber \\
  && + \Sigma \nu \Omega^2 \bra{-\f{3}{2}+\pi\Sigma\f{\varpi^2}{M_\varpi}}^2 \nonumber \\
  && + \f{1}{2\pi \varpi} \f{\p\dot{M}_{\mr{e}}}{\p\varpi} e_{\mr{th,e}}
  - \Lambda_{\mr{rad}}-\Lambda_{\mr{chem}} \h , \label{eq:energy}
\end{eqnarray}
where $e_{\mr{th}}$ is the specific thermal energy of the disk gas
\begin{eqnarray}
 e_{\mr{th}} = \f{1}{\gamma_\mr{ad}-1}\f{k_{B}T}{\mu m_{\chH}} \h ,
\end{eqnarray}
$e_{\mr{th,e}}$ is that of the infalling gas from the envelope, and $\gamma_\mr{ad}$ is the specific heat ratio.
On the right-hand side, the first term represents the advection, the second the viscous heating, the third the energy influx by the mass supply, the fourth the radiative cooling, and the fifth the cooling by chemical reactions such as hydrogen ionization and dissociation.
For simplicity, we evaluate $e_{\mr{th,e}}$ assuming that the gas infalling onto the disk is fully molecular with the temperature $T=1000$~K, and we neglect the shock heating near the disk surface.
We confirmed that the resultant disk structure was almost independent of choosing the different temperatures in the range of $500~\text{--}~5000$~K. We evaluate the radiative cooling term $\Lambda_{\mr{rad}}$ separately in the regions with $T<10^4$~K and $T>10^4$~K, because the disk vertical structure qualitatively changes across the boundary temperature $\sim 10^4$~K \citep[e.g.,][]{Matsukoba_et_al_2019}: the optical depth integrated over the disk scale height is $\lesssim 1$ for $T < 10^4$~K and $\gg 1$ for $T > 10^4$~K.
In the optically thin regime with $T<10^4$~K, $\Lambda_{\mr{rad}}$ is the summation of the continuum cooling $\Lambda_{\mr{cont}}$ and the line cooling by molecular hydrogen $\Lambda_{\chHt}$, which are evaluated in almost the same way as in \citet{Matsukoba_et_al_2019}.
Regarding $\chHt$ collisional induced cooling, we use the formula by \citet{Ripamonti_and_Abel_2004} for $T<500$ K, since the formula by \citet{Matsukoba_et_al_2019} is valid only for $500$ K $< T < 20000$ K.
For the optically thick regime with $T>10^4$~K, on the other hand, we use the diffusion approximation of the thermal continuum emission \citep{Hubeny_1990}:
\begin{eqnarray}
  \Lambda_{\mr{rad}}
  = \f{2}{\Sigma}\sigma_{\mr{SB}}T_{\mr{eff}}^4
  = \f{2}{\Sigma} \f{8\sigma_{\mr{SB}} T^4}{3\tau_{\mr{R}}} \h ,
  \label{eq:Lambda_rad_diff}
\end{eqnarray}
where $\sigma_{\mr{SB}}$ is the Stefan–Boltzmann constant, $T_{\mr{eff}}$ is the effective temperature of the disk surface, and $\tau_{\mr{R}}$ is the optical depth evaluated with the Rosseland mean opacity $\kappa_{\mr{R}}$ as $\tau_{\mr{R}}=\kappa_{\mr{R}}\Sigma/2$.
We use the opacity table provided by the Opacity Project (see \citealt{Seaton_et_al_1994}).
\subsubsection{Chemical Networks}
As for chemical reactions, we take into account the following 5 chemical species, $\chH, \chHt, \chHp, \chHm, \che$.
We compute the abundances of these species using different methods depending on the number density of hydrogen nuclei $n_{\mr{H}}$.
We solve the non-equilibrium rate equations for $\chH,\chHt,\chHp,\che$, assuming the equilibrium abundance only for $\chHm$, when $n_{\mr{H}}<10^{16} \mr{cm}^{-3}$ \citep{Abel_et_al_1997,Matsukoba_et_al_2019}.
We solve the Saha equations when $n_{\mr{H}}>10^{16} \mr{cm}^{-3}$, for which the chemical equilibrium is achieved. The chemical cooling rate is given by
\begin{eqnarray}
 \Lambda_{\mr{chem}} = \bra{\chi_{\chH}\f{dy(\chHp)}{dt}-\chi_{\chHt}\f{dy(\chHt)}{dt}}\f{n_{\chH}}{\rho} \h ,
\end{eqnarray}
where $\chi_{\chH}=13.6$~eV is the ionization energy of hydrogen atom, $\chi_{\chHt}=4.48$~eV is the dissociation energy of molecular hydrogen and $y(\mr{A}) \equiv n(\mr{A})/n_{\mr{H}}$ is a number fraction of species A to hydrogen nuclei.
\subsubsection{Difference from Previous Steady Models}
In previous steady models, the terms including the time derivatives and the mass supply from the envelope in governing equations are neglected.
As a result, the accretion rate, which is often assumed to be equal to the total mass supply rate from the envelope, is a constant throughout the disk.
To calculate the steady disk structure, it is required to give the central stellar mass and disk accretion rate by hand.
In our non-steady model, in contrast, we can determine the accretion rate with $\Sigma$, $T$, and $\dot{M}_e$ at each radius from Equation \eqref{eq:Mdot_d} and can differ from the mass supply rate.
The mass growth history of the central star is determined by the accretion rate at the inner boundary, not by the mass supply rate from the envelope.
\subsection{Overall Procedures and Cases Considered} \label{sec:num_model}
In summary, we solve Equations \eqref{eq:continuity} and \eqref{eq:energy} using the finite volume method to calculate the time evolution of the surface density $\Sigma$ and the midplane temperature $T$. In Equation \eqref{eq:continuity}, we evaluate $\p\dot{M}_{\mr{e}}/\p\varpi$ using Equation \eqref{eq:Mdot_e_varpi} and $\dot{M}_{\mr{d}}$ at the cell boundaries using Equation \eqref{eq:Mdot_d}.
Furthermore, we simultaneously calculate the chemical abundances. We assume a disk forms in the computational domain from $0.1$~AU to $10^5$~AU, which is divided into 480 logarithmically equally-spaced cells. As boundary conditions, we impose the zero-torque condition at the inner boundary and the zero-flux condition at the outer boundary.
We assume that the central protostar immediately accretes the gas flowing in across the inner boundary.
As initial conditions, we only set the protostar at the center without the disk.
The initial stellar mass is set as $10^{-2}~M_\odot$ for models presented in Section \ref{sec:result}, and $0.27~M_\odot$ for comparisons with a 3D simulation in Section \ref{sec:comparison}.
We continue computational runs by the time when the stellar mass reaches $10~M_\odot$ except for the case of the low mass supply rate, i.e., of $K=0.1~K_\mr{fid}$.
In this case, we terminate the run when the stellar mass reaches $7~M_\odot$ to avoid following long-term computation.
This does not affect our conclusions.
We do not follow the evolution after the stellar mass exceeds $10~M_\odot$ in which the protostellar radiative feedback becomes significant \citep[e.g.,][]{Stacy_et_al_2012,McKee_and_Tan_2008, Hosokawa_et_al_2011}.
The evolution under the influence of radiative feedback is out of the scope of the current work.
\par
In this paper, using the model described above, we perform the calculations for six cases in total. In Section~\ref{sec:timeevol}, we show the results for the fiducial case, where $K=K_{\mr{fid}}$ and $f_{\mr{Kep}}=0.5$. In Section \ref{sec:acc}, we study the effects of varying the mass supply rate by setting $K=0.1~K_{\mr{fid}}$ and $10~K_{\mr{fid}}$.
In Section \ref{sec:rot}, we investigate the effects of varying the envelope rotation with $f_{\mr{Kep}}=0.3$ and $0.8$. In Section \ref{sec:comparison}, we present the case tuned for a comparison with the 3D hydrodynamics simulation.
\section{RESULT} \label{sec:result}
\begin{figure*}[htbp]
\begin{center}
  \includegraphics[width=\linewidth]{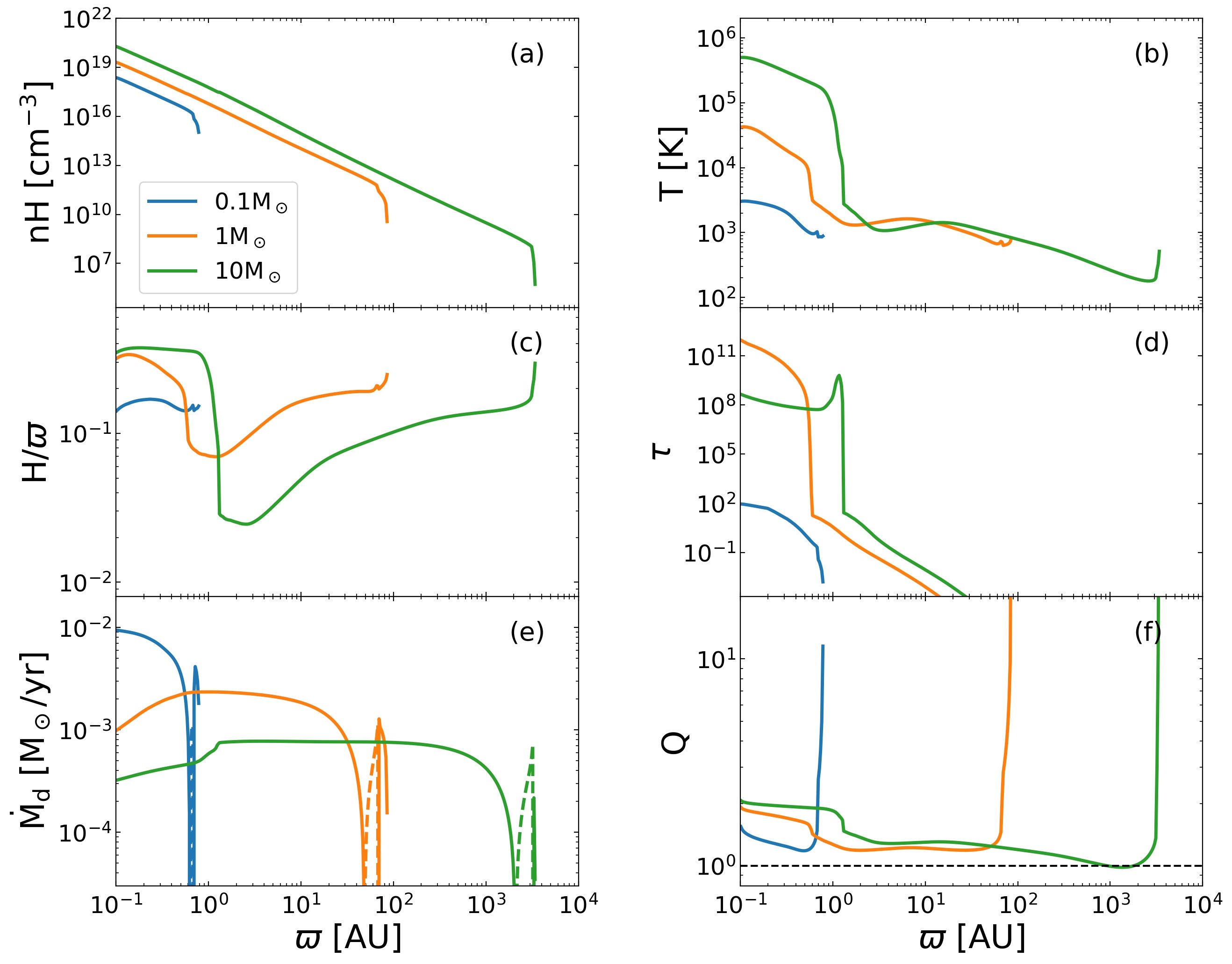}
  \caption{Temporal evolution of the disk structure in the fiducial model, where $K=K_{\mr{fid}}$ and $f_{\mr{Kep}}=0.5$. The line colors represent the different epochs when the stellar mass is 0.1~$M_\odot$ (blue), 1~$M_\odot$ (orange), and 10~$M_\odot$ (green).
  Panels show the radial distributions of (a) gas number density, (b) temperature, (c) disk aspect ratio, (d) optical depth in the vertical direction, (e) accretion rate through the disk, and (f) Toomre $Q$ parameter. In panel (e), the solid and dashed lines correspond to the inward and outward flow, respectively. The horizontal line in panel (f) denotes $Q = 1$.}
  \label{fig:Normal_DiskStructure}
\end{center}
\end{figure*}
\begin{figure}[htbp]
  \begin{center}
    \includegraphics[clip,width=\linewidth]{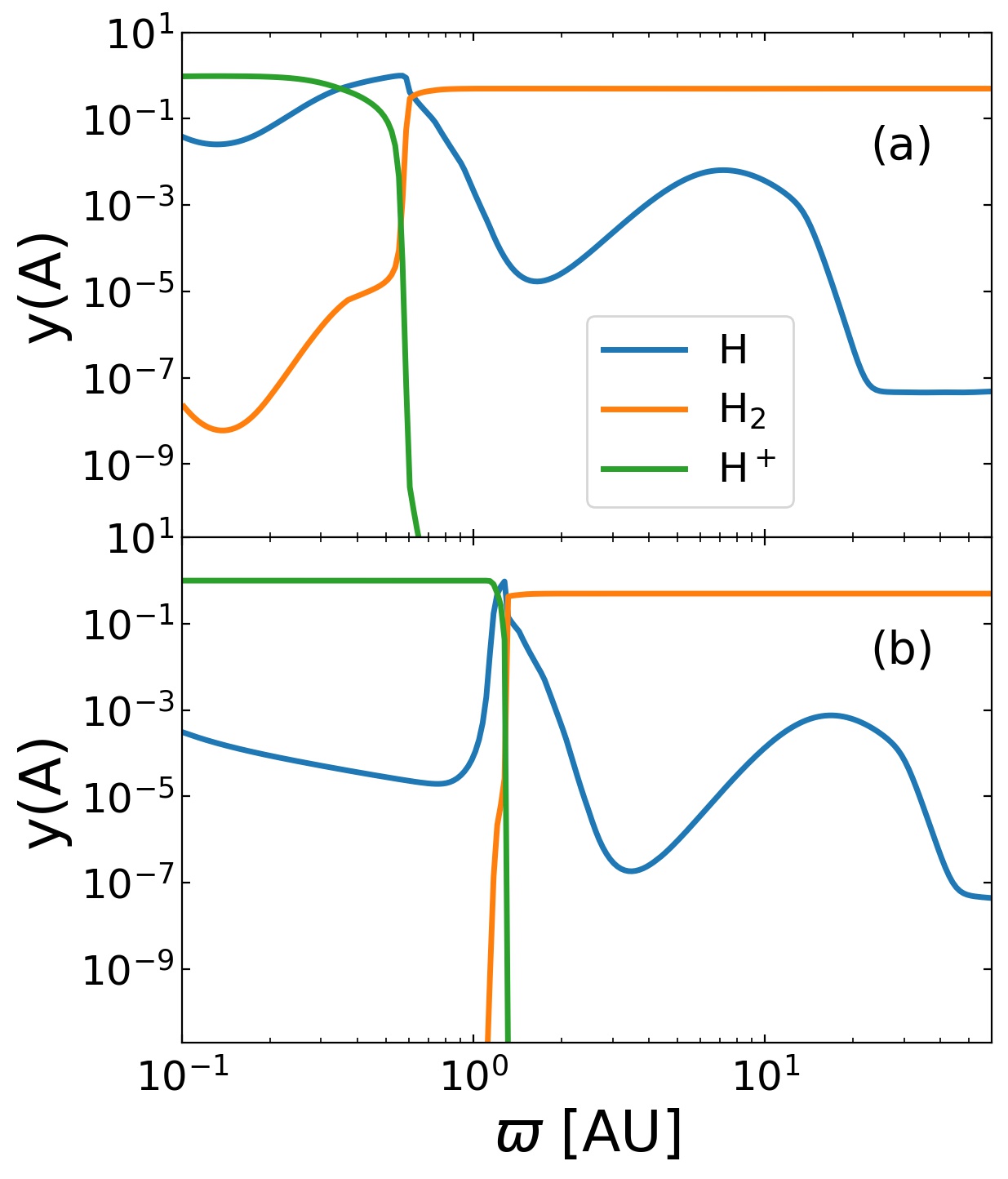}
    \caption{The chemical structure within the disk in the fiducial case for $\varpi<60$~AU. The panels (a) and (b) show the snapshots at the different epochs when $M_*=1~M_\odot$ and $10~M_\odot$.
    The different line colours represent the number fractions of $\chH$ (blue), $\chHt$ (orange), and $\chHp$ (green).}
    \label{fig:Normal_Chem}
  \end{center}
\end{figure}
\begin{figure}[htbp]
  \begin{center}
    \includegraphics[clip,width=\linewidth]{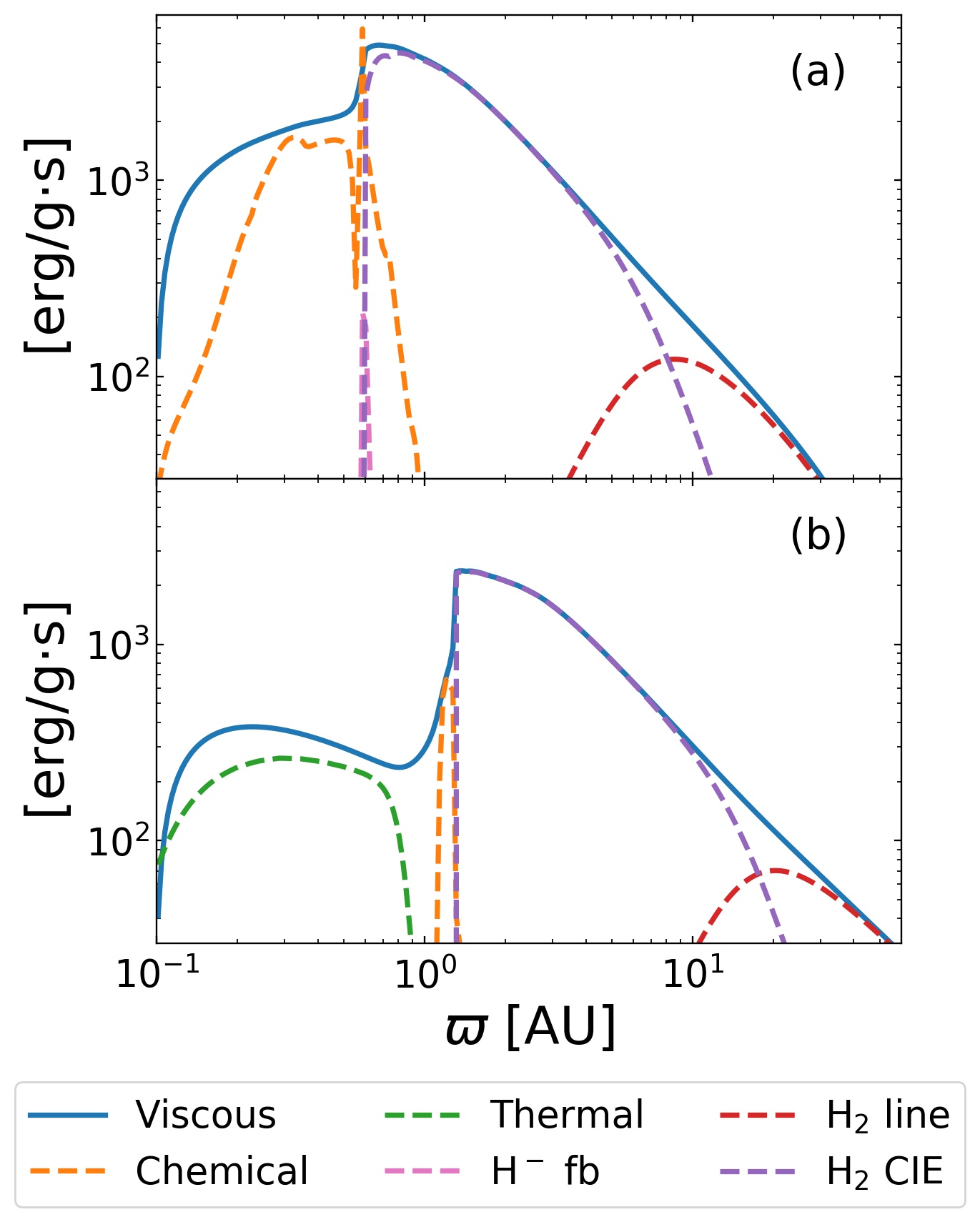}
    \caption{Radial distributions of heating and cooling rates for $\varpi<60$~AU in the fiducial case. The panels (a) and (b) show the snapshots at the same epochs as in Figure~\ref{fig:Normal_Chem}, when $M_*=1~M_\odot$ and $10~M_\odot$. Solid and dashed lines represent heating and cooling processes, respectively. The line colours represent the individual processes as follows: the viscous heating (blue),
    the endotherm by chemical reactions (orange),
    the radiative cooling via the
    $\chHt$ line emission (red),
    $\chHt$ collisional induced emission (purple),
    $\chHm$ free-bound emission (magenta),
    and thermal continuum emission (green).
    }
    \label{fig:Normal_HeatCool}
  \end{center}
\end{figure}
\begin{figure}[htbp]
  \begin{center}
    \includegraphics[clip,width=\linewidth]{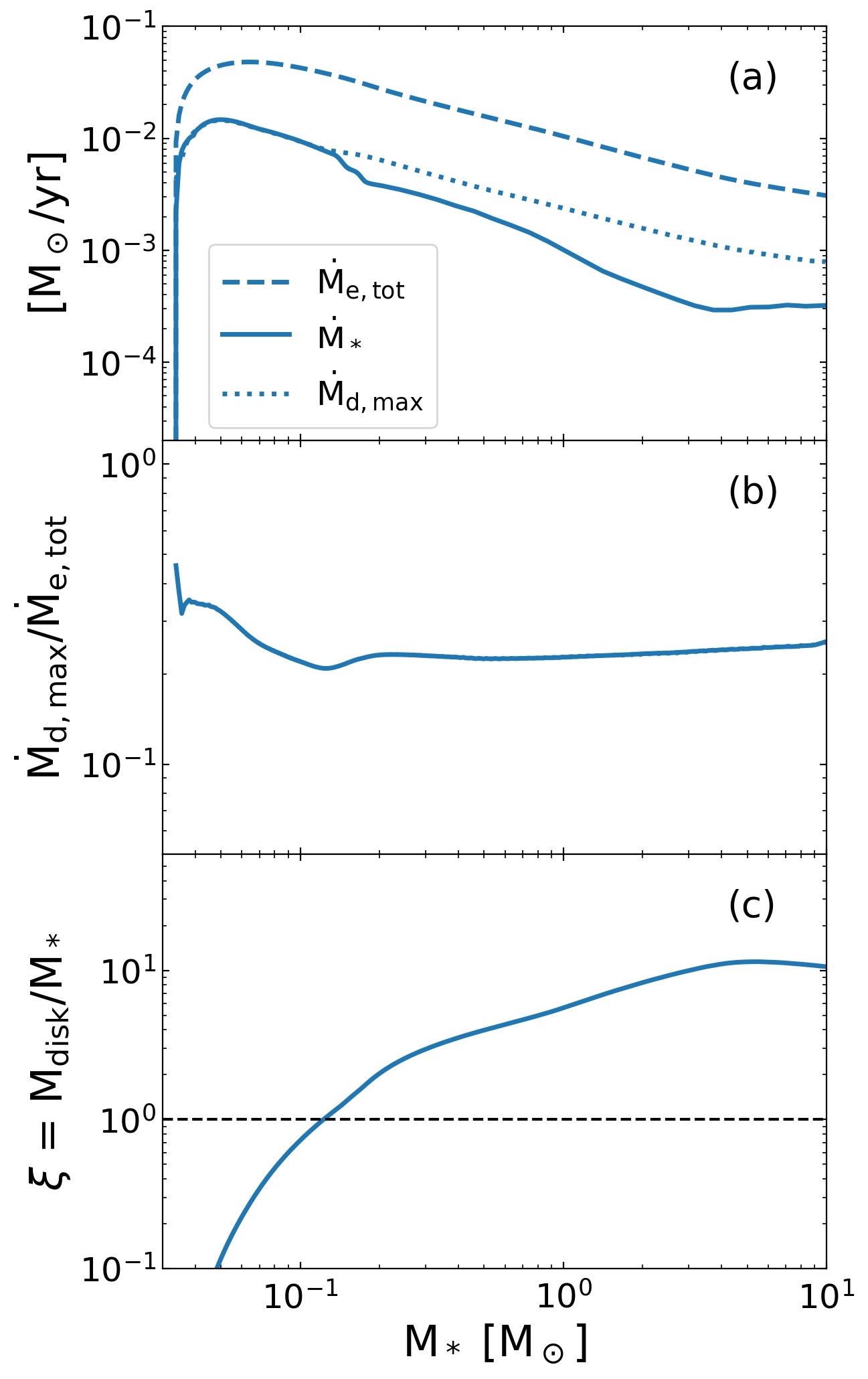}
    \caption{Evolution of the mass flux balance through the disk and disk mass growth history against the stellar mass $M_*$ in the fiducial case. The panel (a) shows the evolution of the total mass supply rate from the envelope onto the disk $\dot{M}_{\mr{e,tot}}$ (dashed line), the accretion rate from the disk onto the star $\dot{M}_*$ (solid line) , and the radial maximum of the disk accretion rate $\dot{M}_{\mr{d,max}}$ (dotted line, also see text). The panel (b) shows the evolution of the ratio of $\dot{M}_\mr{d,max}$ to $\dot{M}_\mr{e,tot}$.
    The panel (c) shows the evolution of the disk-to-star mass ratio $\xi \equiv M_{\mr{disk}}/M_*$ and the horizontal line denotes unity.}
    \label{fig:Normal_TimeEvol}
  \end{center}
\end{figure}
\begin{figure*}[htbp]
  \begin{center}
    \includegraphics[clip,width=\linewidth]{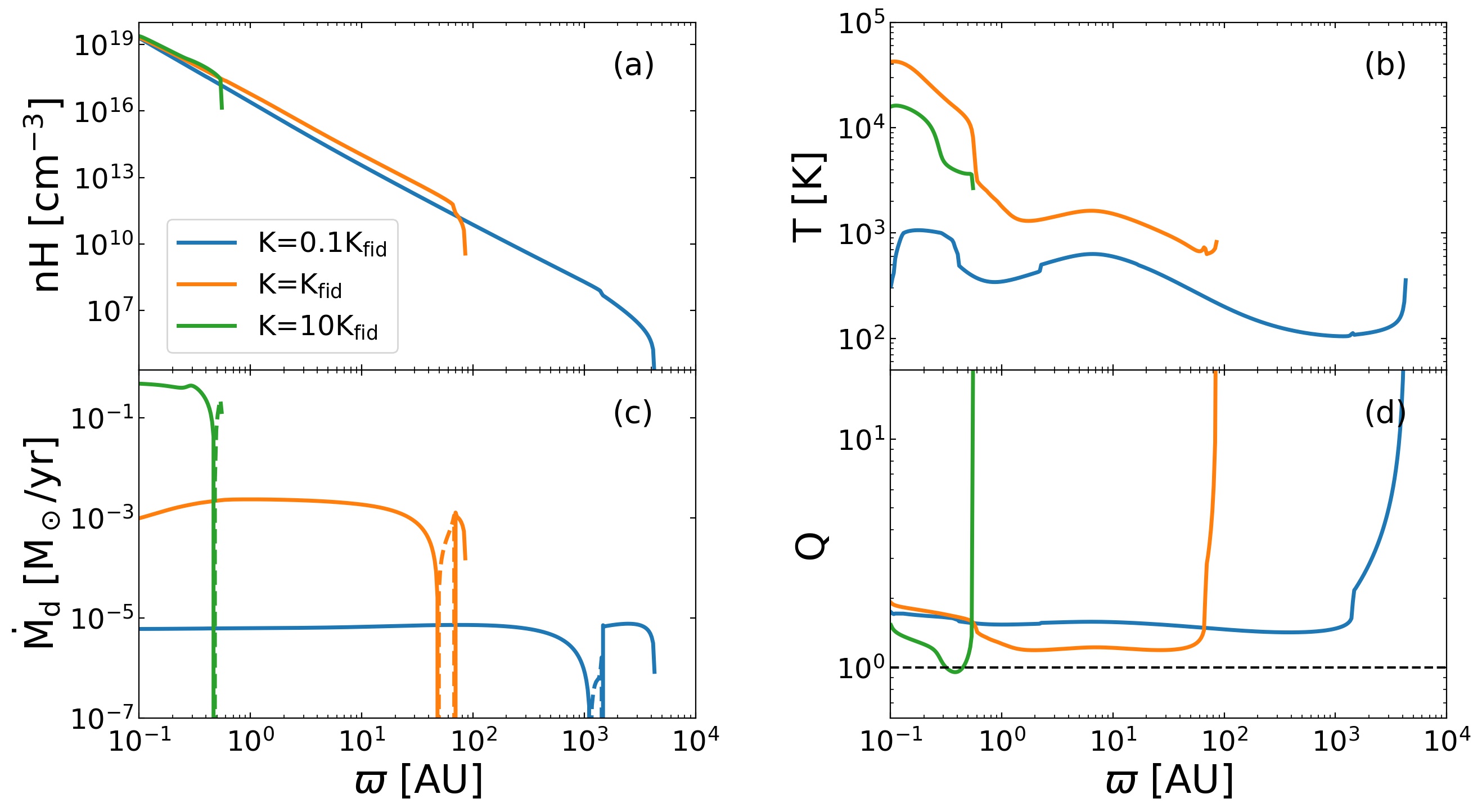}
    \caption{Variations of the disk structure with different mass supply rates from the envelope.
    The blue, orange, and green lines represent the cases with $K=0.1~K_{\mr{fid}}$, $K_{\mr{fid}}$, and $10~K_{\mr{fid}}$, respectively. The same rotation parameter $f_{\mr{Kep}}=0.5$ is assumed for these cases. Plotted are the snapshots when the stellar mass is $1~M_\odot$ for each case.
    Panels show the radial distributions of (a) temperature, (b) number density, (c) accretion rate through the disk, and (d) Toomre $Q$ parameter. In panel (c), the solid and dashed lines correspond to the inward and outward flow, respectively. The horizontal line in panel (d) denotes $Q=1$.
    }
    \label{fig:Acc_DiskStructure}
  \end{center}
\end{figure*}
\begin{figure}[htbp]
  \begin{center}
    \includegraphics[clip,width=\linewidth]{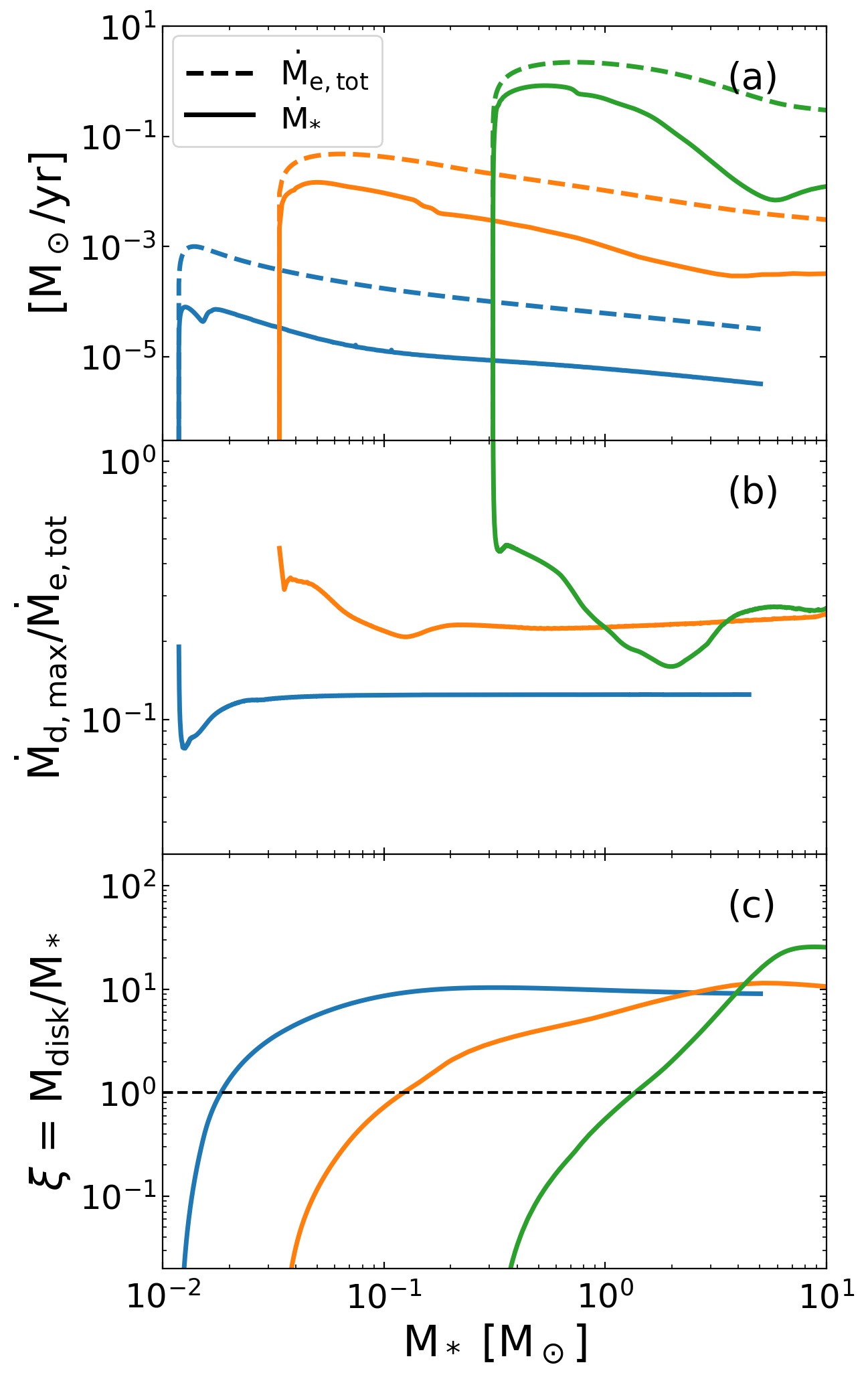}
    \caption{
    Almost the same as Figure~\ref{fig:Normal_TimeEvol} but for the cases with different mass supply rates from the envelope. The line colours represent the same cases as in Figure~\ref{fig:Acc_DiskStructure}:  $K=0.1~K_{\mr{fid}}$ (blue), $K_{\mr{fid}}$ (orange), $10~K_{\mr{fid}}$ (green).
    In the panel (a), we do not show the radial maximum of the disk accretion rate $\dot{M}_{\mr{d,max}}$ for clarity, unlike in Figure~\ref{fig:Normal_TimeEvol}.
}
    \label{fig:Acc_TimeEvol}
  \end{center}
\end{figure}
\begin{figure*}[htbp]
  \begin{center}
    \includegraphics[clip,width=\linewidth]{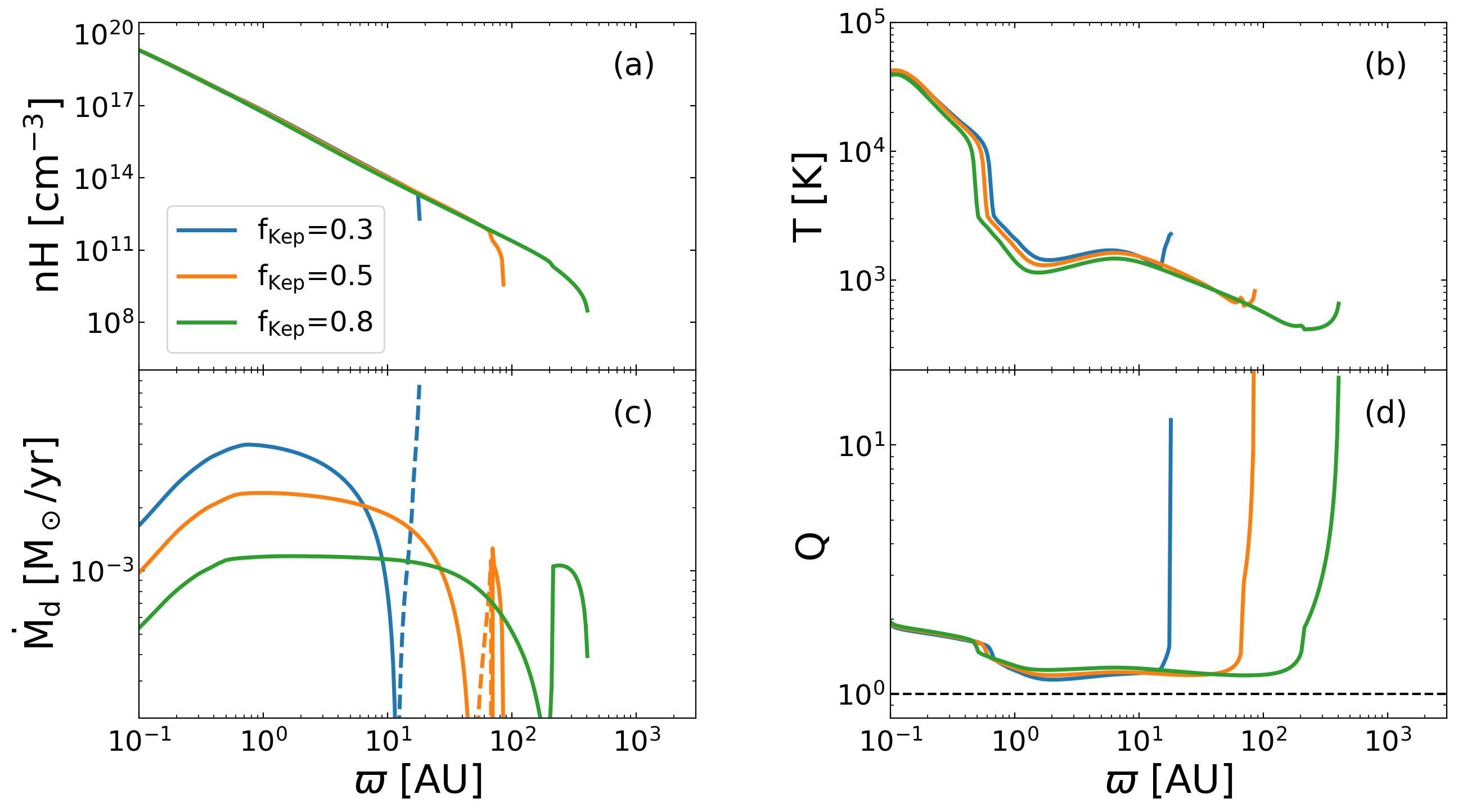}
    \caption{Same as Figure \ref{fig:Acc_DiskStructure} but for the cases with the different rotation degree of the envelope, i.e., with $f_{\mr{Kep}}=0.3$ (blue), 0.5 (orange), and 0.8 (green).
    The same parameter $K=K_{\mr{fid}}$ (see Equations~\ref{eq:env_rho} and \ref{eq:env_v}) is assumed for these cases. Plotted is the snapshot when $M_*=1~M_\odot$ in each case.
    }
    \label{fig:Rot_DiskStructure}
  \end{center}
\end{figure*}
\begin{figure}[htbp]
  \begin{center}
    \includegraphics[clip,width=\linewidth]{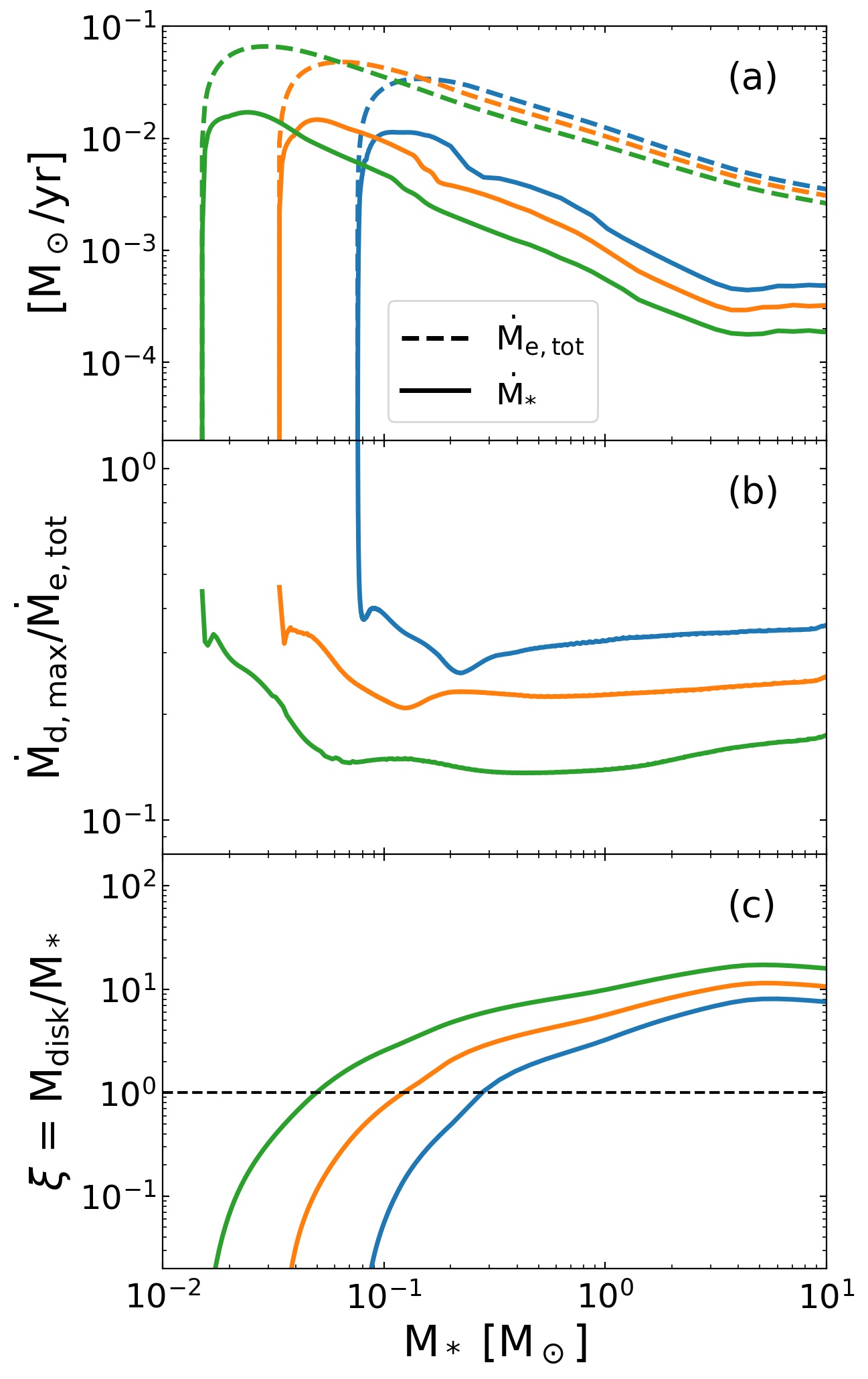}
    \caption{Same as Figure \ref{fig:Acc_TimeEvol} but for the same cases as presented in Figure~\ref{fig:Rot_DiskStructure}, i.e., cases with $f_{\mr{Kep}}=0.3$ (blue), 0.5 (orange), and 0.8 (green).
    }
    \label{fig:Rot_TimeEvol}
  \end{center}
\end{figure}
\subsection{Evolution of Disk in the Fiducial Model} \label{sec:timeevol}
\subsubsection{Disk Structure} \label{subsubsec:model_calculation}
In this section, we overview the fiducial case, where $K=K_{\mr{fid}}$ and $f_{\mr{Kep}}=0.5$. In Figure~\ref{fig:Normal_DiskStructure}, we show the disk structure at the different epochs when $M_*=0.1~M_\odot$, $1~M_\odot$, and $10~M_\odot$. To understand the evolution in Figure~\ref{fig:Normal_DiskStructure}, we also show radial distributions of chemical abundances and effective heating and cooling rates for $\varpi < 60$~AU when $M_*=1~M_\odot$ and $10~M_\odot$ in Figures~\ref{fig:Normal_Chem} and \ref{fig:Normal_HeatCool}.
\par
Figure~\ref{fig:Normal_DiskStructure} overall shows that the disk gradually spreads as the stellar mass increases. The disk outer edge is located at $\sim 1$~AU when $M_* = 0.1~M_\odot$ and $\sim 10^3$~AU when $M_* = 10~M_\odot$.
As shown in Figure~\ref{fig:Normal_DiskStructure}(f), Toomre $Q$ parameter is always regulated to take nearly constant values at $Q~\sim~1$ throughout the disk, except the outermost part. This is a well-known feature of the self-gravitating disk, as also shown by the previous steady 1D models \citep{Tanaka_and_Omukai_2014,Matsukoba_et_al_2019}.
\par 
Figure~\ref{fig:Normal_DiskStructure}(a) shows that the density on the disk midplane rapidly decreases with increasing the radius $\varpi$, which is approximately fitted by the power-law function $n_\chH~\propto~\varpi^{-3}$. This is well interpreted with the following analytical argument.
Since the disk is marginally stable with $Q~\sim~1$, Equation \eqref{eq:Q_def} provides
\begin{eqnarray}
  \Sigma \sim \f{c_s \Omega}{\pi G} \label{eq:Q_to_Sigma} \h.
\end{eqnarray}
Substituting this into Equation (\ref{eq:iso-Sigmarho}), we obtain
\begin{eqnarray}
  \rho_0 = \f{\Sigma}{\sqrt{2\pi}H} \sim \f{\Omega^2}{\sqrt{2\pi^3}G} \propto \varpi^{-3} \h . \label{eq:rho_marginally}
\end{eqnarray}
\par 
As shown in Figure~\ref{fig:Normal_DiskStructure}(b), the midplane temperature is several thousand kelvin throughout the disk when $M_*=0.1~M_\odot$. However, the later snapshots show that the temperature dramatically rises by orders of magnitude only for $\varpi \lesssim 1$~AU, whereas it remains at $T \sim 100~\text{--}~1000$~K in the outer part. Figure~\ref{fig:Normal_Chem} indicates that the cooler outer part is almost fully molecular, and Figure~\ref{fig:Normal_HeatCool} shows that the viscous heating is balanced with H$_2$ molecular cooling via optically thin continuum and line emission (see also Figure~\ref{fig:Normal_DiskStructure}d). On the other hand, Figure~\ref{fig:Normal_Chem} also shows that in the hot inner part hydrogen molecules are almost completely dissociated.
The hot inner part emerges because, as shown in Figure~\ref{fig:Normal_DiskStructure}(d), the optical depth over the disk scale height $\tau_{\mr R} = \kappa_{\mr R} \Sigma/2$ dramatically increases once the temperature exceeds $\sim 3000$~K. This is caused by the contribution of bound-free absorption of $\chHm$ to the opacity $\kappa_{\mr R}$ \citep{Mayer_and_Duschl_2005}. Since the radiative cooling becomes less efficient as $\tau_{\mr R}$ increases, the temperature increases owing to the viscous heating.
With increasing temperature, hydrogen molecules are dissociated and atoms are ionized due to efficient collisions, which absorb the gas thermal energy. However, the endothermic reactions become unavailable from inside out with hydrogen fully ionized (Figures~\ref{fig:Normal_Chem}a and \ref{fig:Normal_HeatCool}a\footnote{The viscous heating decreases rapidly near the inner boundary due to the torque-zero boundary condition.}).
Eventually, the temperature rapidly increases with no effective cooling.
Such vigorous heating continues until the temperature reaches $\sim10^5$~K, above which the opacity almost converges to the Thomson scattering value. The radiative cooling becomes effective again and balances with the viscous heating (Figure \ref{fig:Normal_HeatCool}b). We note that, even when $M_* = 10~M_\odot$, the effective temperature at the disk surface is estimated as only $\simeq 4000$~K from Equation \eqref{eq:Lambda_rad_diff}. The ultra-violet radiation coming from the hot inner part is too weak to cause the photoevaporation in the disk outer part.
\par
Figure \ref{fig:Normal_DiskStructure}(c) shows that the evolution of the disk aspect ratio, $H/\varpi$.
We see that $H/\varpi$ gets lower as the stellar mass increases at a fixed radius, except the innermost part of $\varpi \lesssim 1$~AU for $M_* \gtrsim 1~M_\odot$.
The aspect ratio is nearly $0.3$ in that part, which corresponds to the hot optically thick region described above.
However, our thin disk approximation $H/\varpi \ll 1$ is still valid in the outer region with $T \lesssim 10^3$~K, where the fragmentation is expected to occur.
Note that previous 3D simulations do not spatially resolve the hot inner part, but normally mask it by the sink cells or particles. Further studies are necessary to understand the structure of the innermost part of the disk and its role in the primordial star formation.
\par
We show the accretion rates through the disk $\dot{M}_{\mr d}$ in Figure~\ref{fig:Normal_DiskStructure}(e), where the rate averaged over the disk decreases as the stellar mass increases. This occurs in tandem with the decrease of the mass supply rate from the envelope, as seen in the next section.
Figure~\ref{fig:Normal_DiskStructure}(e) also shows that the gas moves inward near the outer edge of the disk. Such inward motion is caused by the direct mass supply from the envelope to the inner radius, which is represented by the second term on the right-hand side of Equation \eqref{eq:Mdot_d}. The gas would move outward in the disk outermost part only through the viscous evolution, as shown by \citet{Lynden_and_Bell_1974}.
Furthermore, we see that $\dot{M}_{\mr d}$ gradually declines with decreasing $\varpi$ in the hot ionized part for $M_* \gtrsim 1~M_\odot$. Such a feature is not observed in previous models, where the rate is assumed to be constant throughout the disk. Figure~\ref{fig:Normal_DiskStructure}(f) shows that $Q$ in the corresponding part is slightly higher than that in the outer part, suggesting that weak gravitational torque causes the decline of $\dot{M}_{\mr d}$.
Since $\dot{M}_{\mr d}$ decreases with decreasing $\varpi$, the mass transported from the outer part gradually accumulates in the hot ionized disk. This effect operates to promote the gravitational torque, increasing the surface density $\Sigma$, over the timescale of $\tau_{\mr{vis}}\sim\varpi^2/\nu$. As described above, however, the temperature rapidly rises because of the viscous heating until the it reaches $\sim10^5$ K, above which the opacity almost converges to the Thomson scattering value.
The heating effect diminishes the gravitational torque over the timescale of $\tau_{\mr{heat}}\sim e_{\mr{th}}/\nu\Omega^2\sim c_s^2/\nu \Omega^2\sim H^2/\nu$.
The outcome of the above competing effects is judged by the timescale ratio $\tau_{\mr{heat}}/\tau_{\mr{vis}}\sim H^2/\varpi^2$, which is smaller than unity (see Figure \ref{fig:Normal_DiskStructure}c).
Therefore, the latter overcomes the former in our case, meaning that the gravitational torque remains inefficient in the hot ionized part until a while after that the temperature settles at $T \simeq 10^5$~K.
\subsubsection{Disk-to-Star Mass Ratio}
\label{sec:mass_ratio}
In Figure~\ref{fig:Normal_TimeEvol}(a), we show the time evolution of the total mass supply rate from the envelope to the disk $\dot{M}_{\mr{e,tot}}$, the accretion rate from the disk to the central star $\dot{M}_*$, and the radial maximum of the disk accretion rate $\dot{M}_{\mr{d,max}}$.
Figure~\ref{fig:Normal_TimeEvol}(b) shows the ratio $\dot{M}_\mr{d,max}/\dot{M}_\mr{e,tot}$.
Regarding the accretion rate onto the star, strictly speaking, there is an additional supply rate directly from the envelope onto the star.
However, this is much smaller than $\dot{M}_*$ except right after the protostar formation, and we only consider $\dot{M}_*$.
We take the time average of $\dot{M}_\mr{d,max}$ over $\Delta\log(M_*) = 0.1$.
We do this to remove numerical oscillations that appear owing to the limited spatial resolution near the boundary between the inner optically-thick and outer optically-thin parts of the disk.
We have confirmed that such an artifact does not affect the overall evolution by performing experimental calculations with an improved resolution.
Note that $\dot{M}_\mr{d,max}$ represents the approximately constant accretion rate in the outer part of the disk, which is larger than that in the hot inner part as seen in Figure~\ref{fig:Normal_DiskStructure}(e).
\par
Figure~\ref{fig:Normal_TimeEvol}(a) shows that $\dot{M}_{\mr{e,tot}}$ gradually decreases as the stellar mass increases, reflecting the structure of the accretion envelope we have assumed.
Moreover, $\dot{M}_*$ and $\dot{M}_\mr{d,max}$ are an order of magnitude smaller than $\dot{M}_{\mr{e,tot}}$ although they are often assumed equal in steady models.
The ratio $\dot{M}_*/\dot{M}_\mr{e,tot}$ fluctuates around $\sim0.1$, and the ratio $\dot{M}_\mr{d,max}/\dot{M}_\mr{e,tot}$ shown in Figure~\ref{fig:Normal_TimeEvol}(b) remains $\simeq0.2$ throughout the evolution.
In other words, the most of the gas supplied from the envelope is not transferred to the star through the disk.
Figure~\ref{fig:Normal_TimeEvol}(c) shows that the disk becomes more massive than the star when it accretes the gas of only $\sim 0.1~M_\odot$.
In the fiducial case, the disk mass reaches 10 times the central stellar mass at the end of the calculation.
The evolution of the disk-to-star mass ratio $\xi$ is written as follows:
\begin{eqnarray}
 \dot{\xi}&=&\f{d}{dt}\bra{\f{M_{\mr{disk}}}{M_*}} \nonumber \\
 &=& \f{M_*\dot{M}_\mr{e,tot}-(M_*+M_\mr{disk})\dot{M}_*}{M_*^2} \h .
 \label{eq:xi_evol}
\end{eqnarray}
Equation \eqref{eq:xi_evol} indicates that $\xi$ ceases to increase, i.e., $\dot{\xi}~\sim~0$, when $\xi \sim (\dot{M}_\mr{e,tot}-\dot{M}_*)/\dot{M}_* \sim 10$, which well agrees with the evolution presented in Figure~\ref{fig:Normal_TimeEvol}(c).
The disk-to-star mass ratio $\xi$ is a key parameter in considering the disk fragmentation, as further discussed in detail in Section \ref{sec:MassiveDiskInstability}.
\par
To understand the evolution of the disk-to-star mass ratio $\xi$, we here develop an analytical model for the accretion rate through the disk.
We suppose that the disk spreads by $d\varpi_{\mr{d}}$ during a time interval $dt$ due to the mass supply from the envelope, where $\varpi_{\mr{d}}$ is the radius of the disk outer edge.
Hereafter, we consider the mass conservation through the annulus whose inner and outer radii are $\varpi_{\mr{d}}$ and $\varpi_{\mr{d}}+d\varpi_{\mr{d}}$.
We define the accretion rate at the inner radius as $\dot{M}_{\mr{d,edge}} \equiv \dot{M}_{\mr d}(\varpi_{\mr d})$, and assume that the total mass supply from the envelope $\dot{M}_{\mr{e,tot}}$ concentrates on the annulus.
Here, $\dot{M}_\mr{d,edge}$ corresponds to $\dot{M}_\mr{d,max}$ in Figure \ref{fig:Normal_TimeEvol}, which denotes the accretion rate in the outer part.
The mass conservation is described as
\begin{eqnarray}
  dM = \dot{M}_{\mr{e,tot}}dt - \dot{M}_{\mr{d,edge}}dt \h \label{eq:mass_conserv},
\end{eqnarray}
where $dM$ is the mass increment of the annulus.
We further assume that the Toomre Q parameter is regulated to take a constant value $Q_{\mr{crit}}$ within the annulus, where $Q_{\mr{crit}}$ is the critical value below which the gravitational torque is effective. Since the surface density $\Sigma$ is given by
\begin{eqnarray}
 \Sigma = \f{c_s\Omega}{\pi G Q_{\mr{crit}}} \h , \label{eq:marginally_Sigma}
\end{eqnarray}
the mass increment $dM$ is also expressed as
\begin{eqnarray}
 dM=2\pi\varpi_{\mr{d}} d\varpi_{\mr{d}} \Sigma = \f{2 \varpi_{\mr{d}} c_s \Omega}{GQ_{\mr{crit}}} d\varpi_{\mr{d}} \h . \label{eq:dM}
\end{eqnarray}
From Equations \eqref{eq:mass_conserv} and \eqref{eq:dM}, we obtain
\begin{eqnarray}
 \f{\dot{M}_{\mr{d,edge}}}{\dot{M}_{\mr{e,tot}}}
 &=& 1 - \f{1}{\dot{M}_{\mr{e,tot}}}\f{dM}{dt} \nonumber \\
 &=& 1 - \f{1}{\dot{M}_{\mr{e,tot}}}\f{2 \varpi_{\mr{d}} c_s \Omega}{GQ_{\mr{crit}}} \f{d\varpi_{\mr{d}}}{dt} \h . \label{eq:Mdot_ratio}
\end{eqnarray}
The second term on the right-hand side makes the discrepancy between $\dot{M}_{\mr{d,edge}}$ and $\dot{M}_{\mr{e,tot}}$.
This term denotes the effect of the disk spread due to the angular momentum of the infalling gas.
In other words, the gas supplied from the envelope is used to extend the outer edge of the disk.
If the disk does not spread or $d\varpi_{\mr{d}}=0$, Equation \eqref{eq:Mdot_ratio} provides $\dot{M}_{\mr{d,edge}}=\dot{M}_{\mr{e,tot}}$.
To rewrite the second term, we approximate the effective equation of state within the disk as
\begin{eqnarray}
 T = f_T \times 200 \bra{\f{K}{K_\mr{fid}}} \bra{\f{n_\chH}{10^4 \mr{cm}^{-3}}}^{0.1} \mr{K} \h , \label{eq:T_edge}
\end{eqnarray}
where $f_T$ is the temperature ratio of the envelope and disk since the envelope temperature is given by Equation \eqref{eq:TE} in the fiducial case and proportional to $K$.
In the fiducial case, Equation \eqref{eq:T_edge} with $f_T=0.4$ well describes the temperature profiles on the disk midplane except the hot inner part.
We can rewrite the second term in Equation \eqref{eq:Mdot_ratio} using Equation \eqref{eq:T_edge} as described in Appendix \ref{app:analytic}, and finally obtain
\begin{eqnarray}
 \f{\dot{M}_{\mr{d,edge}}}{\dot{M}_{\mr{e,tot}}}
 = 1 &-& 0.68 \bra{\f{Q_{\mr{crit}}}{1.0}}^{-21/20} \nonumber \\
 &&  \bra{\f{f_T}{0.4}}^{1/2} \bra{\f{f_{\mr{Kep}}}{0.5}}^{7/10} \h . \label{eq:Mdot_ratio_2}
\end{eqnarray}
This equation suggests that the ratio $\dot{M}_{\mr{d,edge}}/\dot{M}_{\mr{e,tot}}$ is independent of the stellar mass $M_*$ and it remains $\sim0.3$ throughout the evolution in the fiducial case. These features agree with our numerical result presented in Figure~\ref{fig:Normal_TimeEvol}(b), where $\dot{M}_{\mr{d,max}}/\dot{M}_{\mr{e,tot}}$ remains $\sim0.2$.
Note that our formulation is not applicable when Equation~\eqref{eq:Mdot_ratio_2} returns negative values for $\dot{M}_{\mr{d,edge}}/\dot{M}_{\mr{e,tot}}$.
In this case, $\dot{M}_{\mr{d,edge}}$ becomes negative because the mass supply $\dot{M}_{\mr{e,tot}}dt$ is smaller than the increment of the disk mass $dM$ under the assumption that $Q \sim Q_\mr{crit}$, in Equation \eqref{eq:mass_conserv}.
In other words, the mass supply is not enough to retain $Q < Q_\mr{crit}$.
Hence, our assumption that the Toomre Q parameter is regulated to $Q_\mr{crit}$ due to the gravitational torque is no longer valid.
\par
As shown in Figure~\ref{fig:Normal_TimeEvol}(a), the accretion rate onto the star $\dot{M}_*$ is equal to the maximum disk accretion rate $\dot{M}_{\mr{d,max}}$ for $M_*<0.1~M_\odot$.
After that, the inner hot region emerges and $\dot{M}_*$ is slightly lower than $\dot{M}_{\mr{d,max}}$ (see also Figure~\ref{fig:Normal_DiskStructure}e).
Furthermore, $\dot{M}_*$ returns to increase at $M_*\simeq 4~M_\odot$ and gets closer to $\dot{M}_{\mr{d,max}}$.
When $M_*$ reaches $4~M_\odot$, the opacity almost
converges to the Thomson scattering value and the radiative cooling becomes effective. Hence, timescale balance changes into $t_{\mr{heat}}/t_{\mr{vis}} > 1$ and the discrepancy between $\dot{M}_{\mr{d,max}}$ and $\dot{M}_*$ gradually disappears as described in Section \ref{subsubsec:model_calculation}.
\subsection{Effects of Varying Mass Supply Rates}
\label{sec:acc}
In this section, we consider the cases with $K=0.1~K_{\mr{fid}}$, $K_{\mr{fid}}$, and $10~K_{\mr{fid}}$ (see Equations \ref{eq:Larson_rho} and \ref{eq:Larson_v}) to study the effects of varying the mass supply rate from the accretion envelope onto the disk. The envelope rotation is fixed at $f_{\mr{Kep}}=0.5$ for these cases.
In fact, cosmological simulations show that the mass supply rate varies depending on different properties of star-forming clouds \citep[e.g.,][]{Hirano_et_al_2014, Hirano_et_al_2015}. The fiducial case with $K=K_{\mr{fid}}$ provides the supply rate of $\sim 10^{-2}  M_\odot~{\rm yr}^{-1}$, which is typical in the normal primordial star formation.
The supply rate for the case with $K=0.1~K_{\mr{fid}}$ is $\simeq 1/50$ times lower than that in the fiducial case.
Such low rates are typical in the so-called Pop III.2 cases, where the HD cooling is effective during the cloud collapse \citep[e.g.,][]{Yoshida_et_al_2007,Hosokawa_et_al_2012}.
The rate for the case with $K=10~K_{\mr{fid}}$ is $\simeq 50$ times higher than the standard value, and it almost corresponds to the direct collapse case described in the Introduction.
\par
In Figure~\ref{fig:Acc_DiskStructure}, we show the disk structure when $M_*=1~M_\odot$ for these cases. Figure~\ref{fig:Acc_DiskStructure}(c) shows that the mean accretion rate through the disk is lower with the lower $K$, i.e., the lower mass supply rates.
Note that, with the lower $K$, the envelope gas originally located at the outer part falls onto the disk until the stellar mass reaches $1~M_\odot$ since the envelope density is lower.
Then, the disk size is larger owing to the angular momentum conservation.
Figure~\ref{fig:Acc_DiskStructure}(d) shows that $Q \sim 1$ throughout the disk for all the presented cases, though the averaged value is slightly lower with the higher $K$. For the case with $K=10~K_{\mr{fid}}$, where the disk accretion rate is highest, the minimum value of $Q$ is slightly below unity at $\varpi\simeq 0.4$~AU.
The previous steady models also show the same trend that the disk becomes more gravitationally unstable with increasing the disk accretion rate \citep{Matsukoba_et_al_2019}.
Figure~\ref{fig:Acc_DiskStructure}(a) shows that the number density converges to the power-law distribution $\rho\propto\varpi^{-3}$ shown by Equation (\ref{eq:rho_marginally}) for any cases where $Q \sim 1$.
We show the temperature distributions in Figure~\ref{fig:Acc_DiskStructure}(b).
For the case with $K=0.1~K_{\mr{fid}}$, the temperature is less than $10^3$~K throughout the disk since $Q$ is large and the viscous heating is weak. There is no hot ionized part, which emerges once the temperature exceeds $\simeq 3000$~K (see Section \ref{subsubsec:model_calculation}).
\par
Figure~\ref{fig:Acc_TimeEvol}(a) shows the time evolution of the mass supply rate from the envelope $\dot{M}_{\mr{e,tot}}$ and the accretion rate from the disk to the central star $\dot{M}_*$.
In all the cases, $\dot{M}_*$ is an order of magnitude smaller than $\dot{M}_{\mr{e,tot}}$, and the disk-to-star mass ratio $\xi$ exceeds unity when $M_*$ $\lesssim$ $1~M_\odot$ as shown in Figure~\ref{fig:Acc_TimeEvol}(c).
Figure~\ref{fig:Acc_TimeEvol}(b) shows the ratio of the maximum disk accretion rate $\dot{M}_\mr{d,max}$ to the mass supply rate $\dot{M}_\mr{e,tot}$.
We find that, for the case with $K=0.1~K_{\mr{fid}}$, $\dot{M}_{\mr{d,max}}/\dot{M}_{\mr{e,tot}}$ is almost constant throughout the evolution as in the fiducial case and smaller than that in the other cases.
This trend is interpreted from the viewpoint of Equation \eqref{eq:Mdot_ratio_2} as follows.
Figure~\ref{fig:Acc_DiskStructure}(b) shows that the overall disk temperature is lower with the lower value of $K$. The envelope temperature is also lower obeying the effective equation of state $P=K\rho^\gamma$.
Our models, however, generally show that the the decrease of the disk temperature is always milder than that of the disk temperature. As a result, the temperature ratio $f_T$ tends to be larger with the lower $K$, for which Equation \eqref{eq:Mdot_ratio_2} returns the smaller $\dot{M}_{\mr{d,max}}/\dot{M}_{\mr{e,tot}}$.
\subsection{Effects of Varying Rotation of Envelope} \label{sec:rot}
Next, we show the results for the cases with $f_{\mr{Kep}}=0.3$, 0.5, and 0.8 with the fixed $K=K_{\mr{fid}}$ to understand the effects of varying the rotation of the envelope. A large sample of primordial star-forming clouds show some scatter of $f_{\mr{Kep}}$ around the typical value 0.5 \citep[e.g.,][]{Hirano_et_al_2014}.
\par
Figure~\ref{fig:Rot_DiskStructure} shows the disk structure when $M_*=1~M_\odot$ for these cases.
The disk has almost the same structure for all the cases, except that the disk size and the accretion rate through the disk.
The disk size is larger with the higher $f_{\mr{Kep}}$, i.e., with the more rapid rotation of the envelope due to the angular momentum conservation.
As shown in Figure~\ref{fig:Rot_DiskStructure}(c), the accretion rate through the disk $\dot{M}_{\mr d}$ is lower with the higher $f_{\rm{Kep}}$ on average.
This trend is consistent with Equation \eqref{eq:Mdot_ratio_2} as described below.
\par
In Figure~\ref{fig:Rot_TimeEvol}(a), we show the time evolution of the mass supply rate from the envelope $\dot{M}_{\mr{e,tot}}$, and the accretion rate from the disk to the central star $\dot{M}_*$.
While both the rates $\dot{M}_*$ and $\dot{M}_{\mr{e,tot}}$ are lower with the higher $f_{\mr{Kep}}$ at a given stellar mass, $\dot{M}_*$ is generally more than a order of magnitude smaller than $\dot{M}_{\mr{e,tot}}$.
The disk-to-star mass ratio $\xi$ exceeds the unity well before $M_*$ exceeds $1~M_\odot$ for all the cases as presented in Figure~\ref{fig:Rot_TimeEvol}(c).
Figure~\ref{fig:Rot_TimeEvol}(b) shows the ratio $\dot{M}_\mr{d,max}/\dot{M}_{\mr{e,tot}}$ and we find that it is systematically higher with the lower $f_{\mr{Kep}}$ and almost constant throughout the evolution in any case. Equation \eqref{eq:Mdot_ratio_2} well explains all these dependencies on $f_{\mr{Kep}}$.
\par
\citet{Stacy_and_Bromm_2014} report that the evolution of the total stellar and disk masses strongly depends on the spin of star-forming clouds taken from cosmological simulations, which apparently contradicts our results. However, note that they also show a systematic correlation between the mass supply rate from the envelope and the cloud's spin: the lower rate with the higher spin. In this paper, we consider these effects separately. The evolution in our model shows the same dependence on the mass supply rate as in \citet{Stacy_and_Bromm_2014} (see Sec.~\ref{sec:acc}).
\begin{figure*}[htbp]
  \begin{center}
    \includegraphics[width=\linewidth]{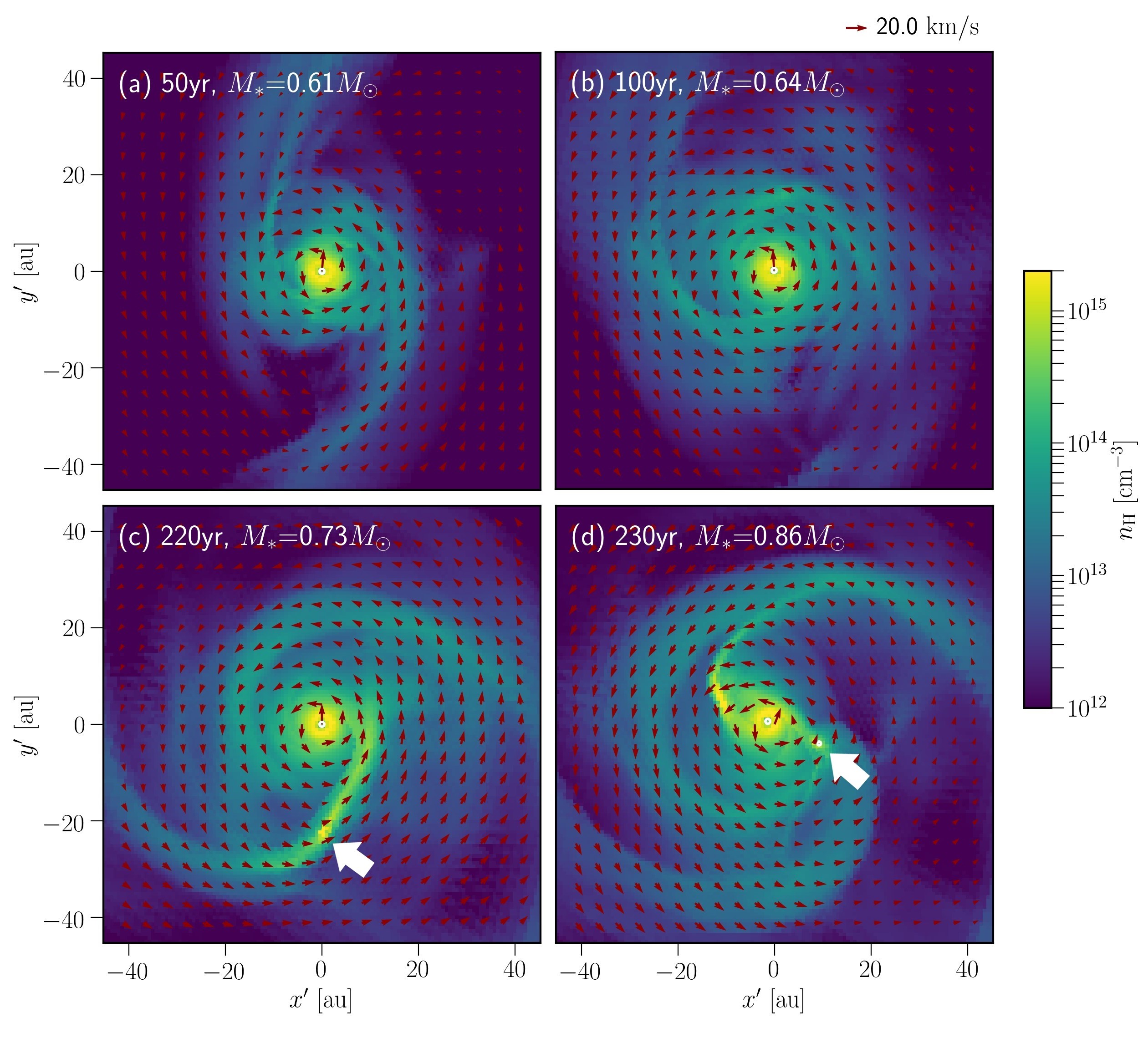}
    \caption{Evolution of the circumstellar disk observed in our 3D simulation. Four panels show the face-on views of the disk at different epochs: (a) 50 yr, (b) 100 yr, (c) 220 yr, and (d) 230 yr after the birth of the protostar. The mass of the central star at each epoch is also presented. In each panel, the color contour and red arrows represent the density and velocity distributions in the disk midplane. The thick white arrows in panels (c) and (d) mark the positions of a fragment.}
    \label{fig:snapshot}
  \end{center}
\end{figure*}
\section{COMPARISON WITH 3D SIMULATION} \label{sec:comparison}
We perform a 3D simulation by using the code SFUMATO-RT developed in \citet{Sugimura_et_al_2020} with some modification.
\citet{Sugimura_et_al_2020} have implemented radiative transfer from primordial protostars to the adaptive mesh refinement code SFUMATO (\citealt{Matsumoto_2007}).
They have also implemented heating and cooling processes and chemical network of primordial gas relevant for the density of $n_{\chH}<10^{13}$ $\mr{cm}^{-3}$.
They take a typical primordial star-forming cloud from the cosmological simulations (\citealt{Hirano_et_al_2014}; \citealt{Hirano_et_al_2015}) as initial condition. They set the sink radius $r_{\mr{sink}}=64$ AU and follow the long-term evolution over $1.2\times10^5$~yr after the first protostar formation, until the final stellar masses are determined by protostellar radiative feedback. In their simulation, the first disk fragmentation occurs at $\sim 1000$~yr after the first protostar formation, when $M_* \simeq 10~M_\odot$.
\par
Previous simulations already show that the disk tends to fragment in the earlier phase with a higher spatial resolution. For example, \citet{Clark_et_al_2011} follow the evolution with a sink radius of $0.5$ AU, reporting that the disk fragments when $M_* \simeq 0.5~M_\odot$.
Furthermore, \citet{Greif_et_al_2012} perform simulations with the higher resolution of $0.05~R_\odot$ and find the disk fragments when $M_*$ is smaller than $0.1~M_\odot$ for all their examined cases.
We here also consider such early fragmentation when $M_* < 1~M_\odot$, for which we have developed 1D non-steady models as presented in Section \ref{sec:result}.
We use the latest version of the code incorporating the cooling processes effective for the higher densities up to $n_{\chH} \sim 10^{16}~\mr{cm}^{-3}$, such as $\chHm$ free-free emission, $\chHt$-$\chHt$ and $\chHt$-
He collisional induced emission, and start the simulation from the same initial condition as in \citet{Sugimura_et_al_2020}. We assume more than hundred times smaller sink radius than in \citet{Sugimura_et_al_2020}, $r_{\mr{sink}}=0.5$~AU.
To reduce the computational cost, we ignore the irradiation from the accreting stars since its effect is negligible compared to compressional and viscous heating in the regime of our simulation \citep{Clark_et_al_2011}.
\par
Figure~\ref{fig:snapshot} shows the simulation snapshots of the number density of hydrogen nuclei from the face-on view at 50, 100, 220, and 230~yr after the first protostar formation.
We see that the disk forms around the central star and finally fragments.
The first fragmentation occurs at $\sim220$~yr after the protostar formation when the stellar mass is $0.73~M_\odot$ (panel c).
White arrows in panels (c) and (d) mark the positions of a fragment. We terminate the simulation at $250$~yr after the protostar formation.
\par
For comparisons, we also follow the disk evolution using the 1D model from the initial condition mimicking the envelope structure observed in the simulation.
We choose the parameters in Equations \eqref{eq:Larson_rho} and \eqref{eq:Larson_v} so that its profiles match the spherically-averaged profile of the simulation result.
We adopt $K=1.07~K_{\mr{fid}}$, $C_1 =6.1\times10^7$, $C_2=-4.5$, and $\gamma=1.1$. Although $C_1$ and $C_2$ are originally constants and not parameters, we adjust $C_2$ to fit the model envelope with the simulation envelope. We locate the inner boundary at 0.1~AU.
\begin{figure}[tbp]
  \begin{center}
     \includegraphics[width=\linewidth]{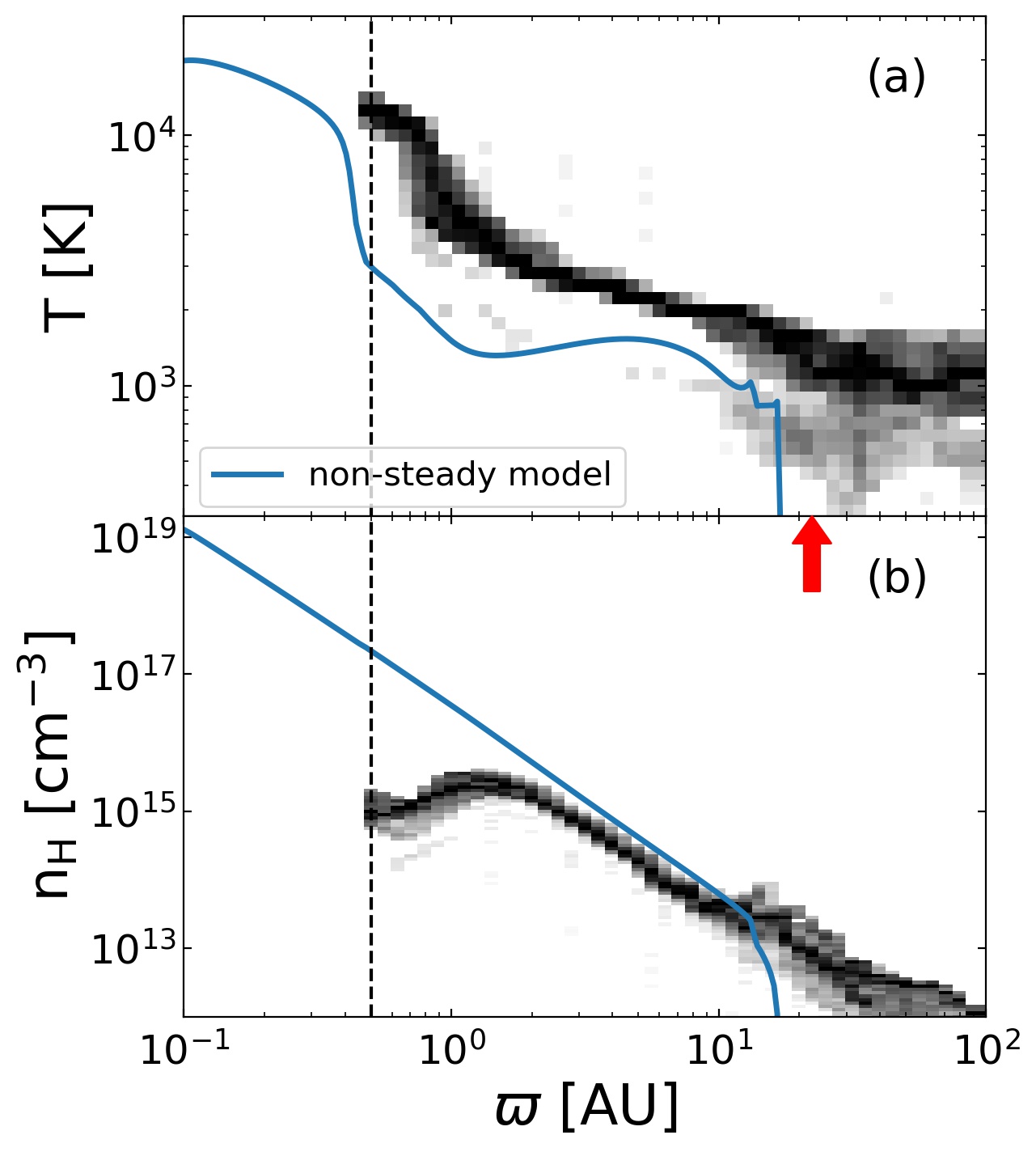}
      \caption{Comparisons of radial distributions of the (a) temperature and (b) number density between the non-steady model and simulation.
      The snapshots are taken at the same time as in Figure~\ref{fig:snapshot}(b), the epoch of 100~yr after the birth of the protostar.
      Blue solid lines represent the 1D non-steady model.
      Black shades show the mass distributions in the simulation data.
      Vertical broken line indicates the sink radius of the simulation, $0.5$~AU. The red arrow in panel (b) marks the estimated disk outer boundary in the simulation (also see text).}
      \label{fig:disk100yr}
  \end{center}
\end{figure}
\par
\subsection{Disk Structure}
Figure~\ref{fig:disk100yr} shows the radial distributions of the temperature and number density at the disk midplane in the model and the simulation at 100 yr after the protostar formation.
While the disk gradually expands with the mass supply from the envelope, its temperature and density profiles are similar during most of our computational time. We thus only consider the disk structure at this epoch as a representative snapshot.
\par
As seen in Figure~\ref{fig:disk100yr}(a), the model and the simulation have a common trend that the temperature is several thousand kelvin in the outer region ($\varpi>0.5$ AU in the model and $\varpi>1$ AU in the simulation) and it rapidly increases toward the central star.
In the simulation data, however, the temperature is higher than that in the model by a factor of $\simeq 2$ overall and it increases monotonically with decreasing $\varpi$.
The same trends are also found in the simulation results reported by \cite{Clark_et_al_2011}.
Figure~\ref{fig:disk100yr}(b) shows that, concerning the number density, they are consistent in the range $\varpi > 2$ AU. The simulation data shows that the number density declines inside $\varpi \simeq 2$ AU owing to the influence of the sink particle.
\par
Although there is no distinct outer boundary of the disk in the simulation, we determine the disk size as follows.
First, we estimate the radius $r(t)$ inside which the gas reaches the disk by the time $t$ using Equation \eqref{eq:t_infall} with the spherically-averaged radial velocity at the moment when the protostar is formed.
Next, with the angular momentum of the envelope $J_\mr{env}(r)$, we define the angular momentum supplied to the disk at the time $t$ as $J_\mr{env}(t)=J_\mr{env}(r(t))$.
Finally, we determine the disk size $\varpi_\mr{d}$ so that $J_\mr{disk} (\varpi_\mr{d}) = J_\mr{env} (t)$.
In Figure~\ref{fig:disk100yr}, the resulting disk outer edge in the simulation is $\simeq 20$ AU, as marked by the red arrow. The difference from the model is less than a factor of two.
Moreover, the simulation data shows the component with the low temperature $T < 10^3$~K in the outer region of $\varpi \gtrsim 10$~AU. This is the residual of the original envelope structure. The similar low-temperature component originally extends to $\varpi \lesssim 10$~AU, but it disappears as the disk spreads radially. This fact also indicates the disk outer edge is located at $\varpi \sim 10$~AU in the simulation data.
\begin{figure}[tbp]
  \begin{center}
    \includegraphics[width=\linewidth]{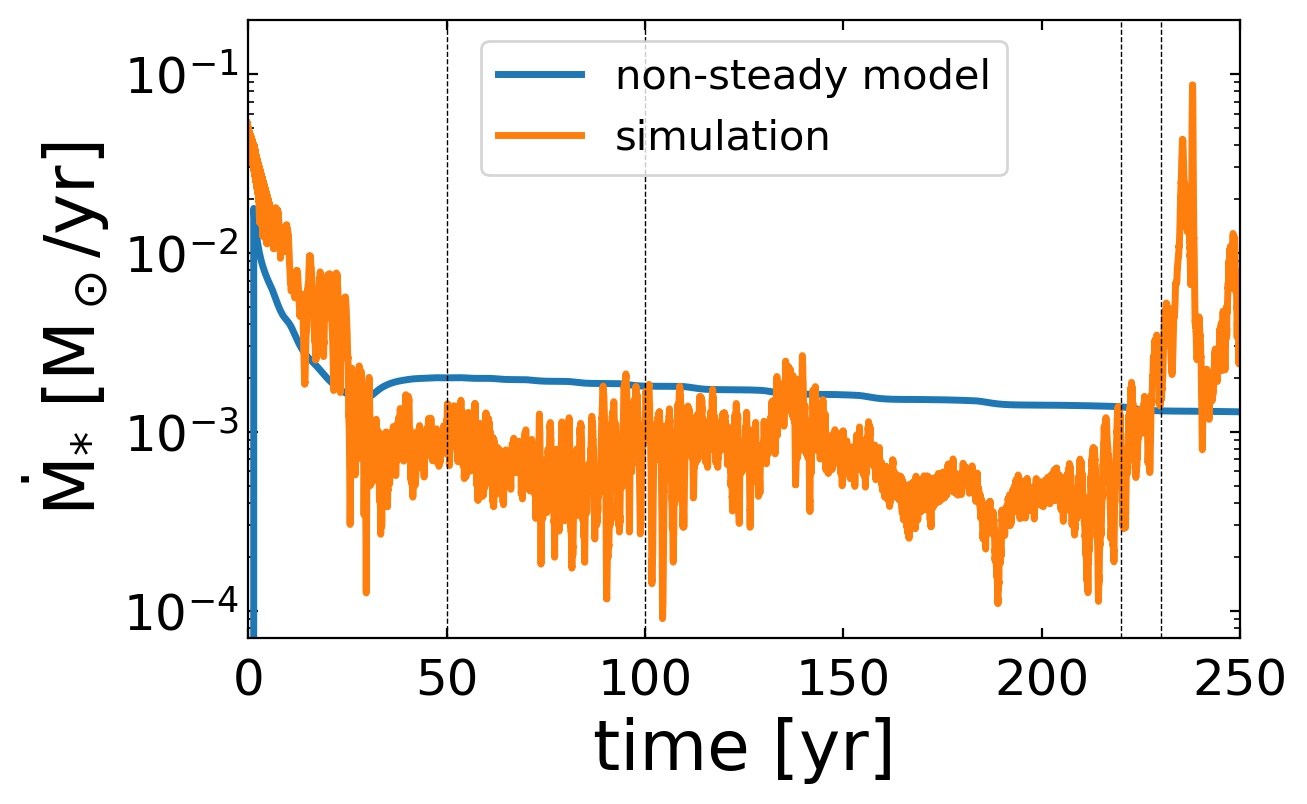}
    \caption{Comparisons between the non-steady model (blue) and simulation (orange), on the mass accretion histories onto the central star. The time for the simulation is measured from the formation of the first sink. Vertical lines indicate the epochs of 50, 100, 220, and 230 yr after the protostar formation, which correspond to the snapshots presented in Figure~\ref{fig:snapshot}.}
    \label{fig:sink_evol_5AU}
  \end{center}
\end{figure}
\begin{figure}[tbp]
  \begin{center}
    \includegraphics[clip,width=\linewidth]{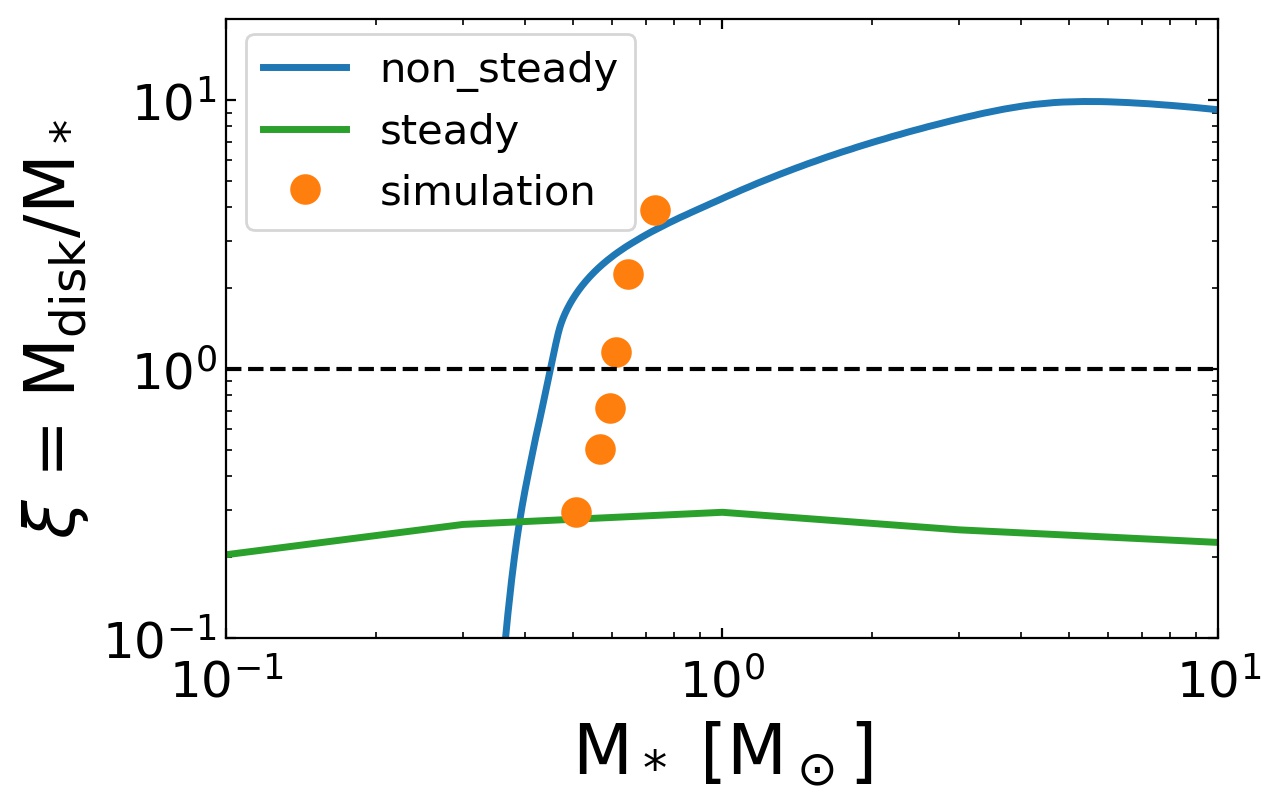}
    \caption{Comparisons between the non-steady model (blue line) and simulation (orange data), on the evolution of the disk-to-star mass ratio $\xi=M_{\mr{disk}}/M_{*}$. In further comparison, the evolution in the 1D steady disk model by \citet{Tanaka_and_Omukai_2014} is also shown by the green line. The horizontal dashed line represents $\xi = 1$, i.e., when the disk and star have the equal masses.}
    \label{fig:xi_compari}
  \end{center}
\end{figure}
\subsection{Accretion Rate and Disk-to-Star Mass Ratio}
In this section, we compare the evolution of the accretion rate onto the star $\dot{M}_*$ and the disk-to-star mass ratio $\xi=M_\mr{disk}/M_*$ in our non-steady model and the simulation.
Regarding $\xi$, we also show the result of the previous steady model for comparison.
\par
Figure~\ref{fig:sink_evol_5AU} shows the mass accretion histories onto the central star in the model and the simulation. The vertical lines denote the elapsed times of 50, 100, 220, and 230 yr after the protostar formation, when we present the snapshots in Figure~\ref{fig:snapshot}.
Whereas the accretion rates in the simulation show the significant time variability, the mean rates in the model are only slightly higher than those in the simulation until 220 yr.
In the simulation, the accretion rate rapidly rises for $t > 220$~yr, leading to the burst-like event at $t \simeq 240$~yr. This is caused by the inward migration of the fragment as seen in Figure~\ref{fig:snapshot}(c) and (d).
\par
Figure~\ref{fig:xi_compari} shows the evolution of the disk-to-star mass ratio $\xi$.
In addition to our non-steady model and 3D simulation, we here also reproduce the steady disk model developed by \citet{Tanaka_and_Omukai_2014} for comparison.
Our non-steady disk model (blue line) shows the rapid increase of $\xi$ in the beginning, and the disk mass exceeds the stellar mass at $M_* \simeq 0.45~M_\odot$.
After that, $\xi$ gradually increases and finally reaches $\sim10$.
This is because the accretion rate onto the star is an order of magnitude smaller than the mass supply rate from the envelope as seen in Section \ref{sec:mass_ratio}.
The simulation (orange dots) also shows that $\xi$ increases rapidly in the beginning. The mass ratio $\xi$ exceeds unity at $M_* \simeq 0.6~M_\odot$, and the disk eventually fragments at $M_* \simeq 0.73~M_\odot$ when $\xi \simeq 3.9$. This implies that in the simulation the gravitational torque in the disk transfers to the protostar only a fraction of the total gas supplied from the envelope, as in the non-steady model.
The green line denotes $\xi$ of the steady model by \citet{Tanaka_and_Omukai_2014}.
They consider the envelope structure with the thermal evolution during the collapse followed by a one-zone model \citep{Omukai_et_al_2005,Hosokawa_and_Omukai_2009}, and suppose the accretion rate through the disk to be equal to that from the envelope, $\dot{M}\sim c_s^3/G$. In their model, the central stellar mass is estimated as the total mass supplied from the envelope by a given epoch, assuming that stellar mass always dominates the disk mass.
In this case, $\xi$ remains almost constant at $0.2\sim 0.3$ throughout the evolution in contrast to the non-steady model and the simulation.
\section{DISCUSSION} \label{sec:discussion}
\subsection{Instability of Massive Disk} \label{sec:MassiveDiskInstability}
\begin{figure}[tbp]
  \begin{center}
    \includegraphics[clip,width=0.95\linewidth]{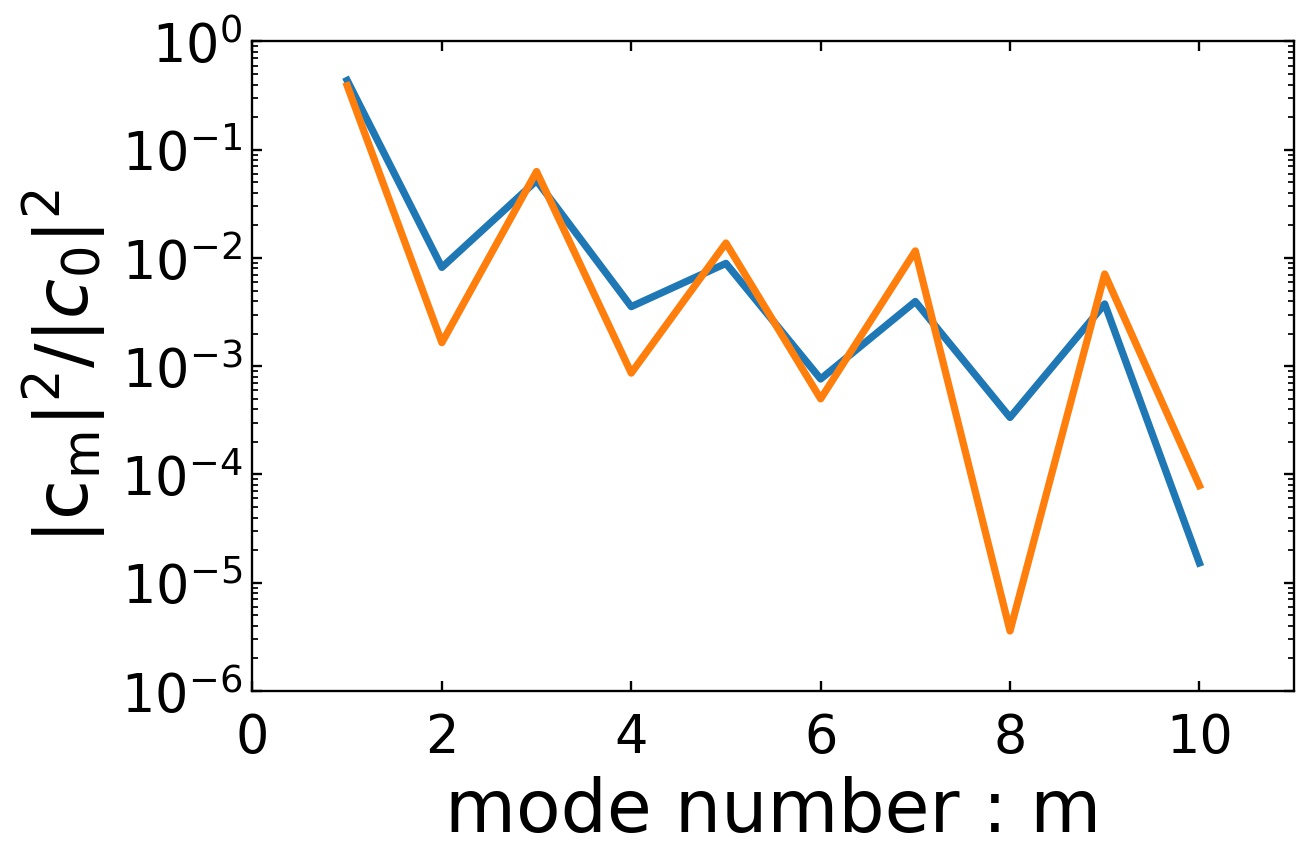}
    \caption{Normalized power $|c_m|^2/|c_0|^2$ in azimuthal modes $m$ in the disk calculated with the simulation data. The blue and orange lines represent the snapshots at the epochs of 50 and 100 yr after the protostar formation, the same as in Figure~\ref{fig:snapshot}(a) and (b).}
    \label{fig:ModePower}
  \end{center}
\end{figure}
We have developed the 1D non-steady model to follow the disk evolution, considering the mass supply from the surrounding accretion envelope. Our model shows that the disk becomes more massive than the central star soon after the protostar formation because the accretion rate through the disk is $\sim 10$ times lower than the mass supply rate from the envelope. We have also confirmed that this trend agrees with the evolution observed in the 3D simulation.
\par
Previous studies show that a specific effect is important in the gravitational stability of the disk much more massive than the central star \citep[e.g.,][]{Kratter_and_Lodato_2016}.
When the disk is massive enough, the protostar no longer stays at the mass center of the system owing to the graviational force exerted from the perturbed disk. It has been known that the perturbation with $m=1$ mode rapidly grows for such a case, even if the Toomre Q value is well above the unity \citep[SLING instability, e.g.,][]{Adams_et_al_1989, Shu_et_al_1990}.
Although only the linear analyses do not predict that the SLING instability necessarily leads to the fragmentation, numerical simulations for the present-day star formation show that the disk fragmentation generally occurs when the disk-to-star mass ratio $\xi$ exceeds unity \citep[][]{Kratter_et_al_2010}.
\par
To investigate whether the SLING instability operates in our 3D simulation, we analyze the disk structure before the fragmentation, following \citet{Krumholz_et_al_2007}. We compute the power $c_{m}$ of the disk spiral arms,
\begin{eqnarray}
  c_m = \f{1}{2\pi} \int_{-\pi}^\pi d\phi \int_0^{\infty} d\varpi e^{im\phi} \varpi \Sigma \h .
\end{eqnarray}
Figure~\ref{fig:ModePower} shows the normalized power $|c_{m}|^2/|c_0|^2$ with the modes of $m = 1~\text{--}~10$ derived for the different epochs whose snapshots are presented in Figure~\ref{fig:snapshot}(a) and (b).
As seen in the figure, the $m=1$ spiral mode is dominant and odd modes are always stronger than even modes. The contrast between the even and odd modes is amplified during the evolution leading to the fragmentation. The same trend is also reported in \citet{Krumholz_et_al_2007}, and it is a possible signature of the SLING instability.
\par
As mentioned above, SLING instability triggers the growth of the perturbation with $m=1$ mode, but it does not necessarily lead to the fragmentation.
While the fragmentation condition is not yet clear, many authors have studied the condition \citep{Gammie_2001,Rice_et_al_2003,Tsukamoto_et_al_2015,Takahashi_et_al_2016_fragmentation,Inoue_and_Yoshida_2018,Inoue_and_Yoshida_2020}.
In particular, \citet{Takahashi_et_al_2016} analytically derive the fragmentation condition of spiral arms forming in circumstellar disks, which agrees with their two-dimensional simulation results. \citet{Inoue_and_Yoshida_2020} modify the condition of \citet{Takahashi_et_al_2016} concerning disk thickness and shearing motion, and apply it for the cosmological simulation performed by \citet{Greif_et_al_2012}. They show that the disk fragmentation in the primordial star formation is also explained in essentially the same framework.
However, it is still unclear what mechanism builds up spiral arms that satisfy their fragmentation condition.
Our results imply that in the primordial star formation the disk becomes massive as the stellar accretion rate is much lower than the mass supply rate from the envelope, resulting in the onset of the SLING instability that leads to the formation of such spiral arms.
\subsection{Resolution Dependence}
\begin{figure}[tbp]
  \begin{center}
    \includegraphics[clip,width=0.95\linewidth]{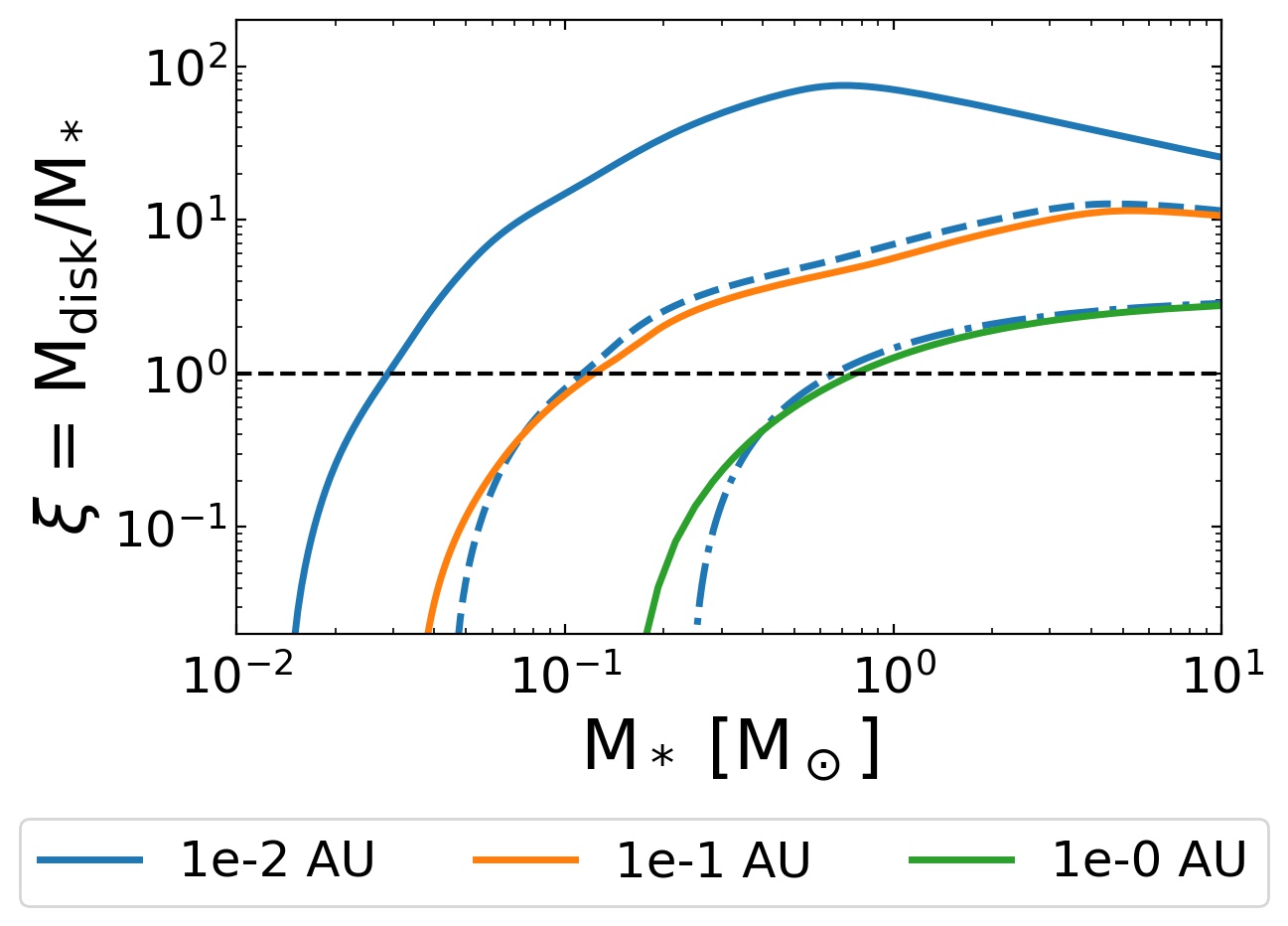}
    \caption{Effects of varying the inner boundary positions in 1D non-steady models. Plotted are the evolution of the disk-to-star mass ratio $\xi = M_{\mr{disk}}/M_*$ with the same envelope structure as in the fiducial model (Section \ref{sec:timeevol}) with different inner boundary radius, 0.01~AU (blue), 0.1~AU (orange) and 1~AU (green). The blue dashed and dot-dashed lines show $\xi$ when we regard the disk mass inside 0.1 AU and 1 AU as the stellar mass using the same inner boundary at $0.01$~AU.}
    \label{fig:xi_resolution}
  \end{center}
\end{figure}
We have shown that, in our 1D disk model, the disk generally becomes more massive than the central star before the stellar mass exceeds 1 $M_\odot$ (see Section \ref{sec:result}).
We here note that the disk-to-star mass ratio $\xi=M_\mr{disk}/M_*$ is highly resolution-dependent. Figure~\ref{fig:xi_resolution} shows the evolution of $\xi$ with the different positions of the inner boundary, 0.01 AU, 0.1 AU, and 1 AU. The same envelope structure as in the fiducial case, where $K = K_{\mr{fid}}$ and $f_{\mr{Kep}}=0.5$, is assumed for these experimental runs.
In the case with the smallest boundary at 0.01 AU (the blue solid line), $\xi$ exceeds the unity early at $M_* \simeq 0.03~M_\odot$. Meanwhile, with the largest boundary at 1 AU (the solid green line), $\xi \simeq 1$ when the star is more massive, at $M_* \simeq 0.7~M_\odot$. Note that the above variation is not due to the different disk evolution caused by the change of the inner radius.
To show this, we consider additional two cases in which the inner boundary is fixed at 0.01 AU but the disk mass inside 0.1 AU (1 AU) is regarded as a part of the central stellar mass.
The blue dashed and dot-dashed lines in Figure~\ref{fig:xi_resolution} represent these cases. We see that these lines almost coincide with the orange and green lines, indicating that the disk evolution is insensitive to the radius of the inner boundary.
\par
In the 1D model, we can choose an arbitrarily small position of the inner boundary. However, its minimum value should be physically limited by the position of the stellar surface. Stellar evolution calculations show that the protostellar radius evolves depending accretion histories \citep[e.g.,][]{Stahler_et_al_1980,Omukai_and_Palla_2001,Omukai_and_Palla_2003,Hosokawa_and_Omukai_2009b}. In the typical cases of primordial star formation, where the accretion rate is $\sim 10^{-3}~\text{--}~10^{-2}$~M$_\odot {\rm yr}^{-1}$, the evolution of the stellar radius is well approximated by
\begin{eqnarray}
 R_*
 \simeq 26~R_\odot
 \left( \frac{M_*}{M_\odot} \right)^{0.27}
 \left( \frac{\dot{M}_*}{10^{-3}~{\mr M}_\odot {\rm yr}^{-1}} \right)^{0.41} \h ,
\end{eqnarray}
in the so-called early adiabatic accretion stage for $M_* \lesssim 10~M_\odot$ \citep[][]{Stahler_et_al_1986}.
Since the above corresponds to $R_* \sim 0.1$~AU, our default choice of $0.1$~AU as the inner boundary radius in Section~\ref{sec:result} is justified.
In Figure~\ref{fig:xi_resolution}, the orange line represents this most realistic case.
\par
On the other hand, 3D simulations are also resolution dependent.
However, it is reasonable to regard that high-resolution simulations by \citet{Greif_et_al_2012} are the most reliable among the previous simulations, since they follow an early evolution of the protostellar accretion even resolving the stellar interior structure.
They show that the disk fragmentation occurs early when the stellar masses are only $0.1~\text{--}~0.3~M_\odot$ for all their examined cosmological mini-halos.
This early fragmentation well matches the implication of our non-steady model that the disk becomes massive so early, leading to the fragmentation, as mentioned in Section \ref{sec:MassiveDiskInstability}.
\par
The above consideration strongly suggests that the early disk fragmentation inevitably occurs.
Whereas this apparently makes the 1D disk models inapplicable to the later evolution, it is not necessarily the case. Extending 1D models should be still helpful for the cases where the disk fragmentation is suppressed or delayed by some processes, e.g., mass and angular momentum transfer by magnetic forces \citep[e.g.,][]{Machida_and_Doi_13}.  It is also possible that a binary system surrounded by a global disk emerges after the fragmentation event \citep[e.g.,][]{Sugimura_et_al_2020}. Semi-analytic modeling of such circumbinary disks is to be studied for understanding the mass growth of binary systems.
\section{SUMMARY AND CONCLUSION} \label{sec:summary}
We have developed a 1D semi-analytical model of the disk forming around a primordial protostar, taking the mass supply from an envelope onto the disk into account. We do not assume the steady accretion through the disk, i.e., the mass transfer rates through the disk is neither constant nor equal to the mass supply rate from the envelope.
We have systematically studied how the disk evolution varies with different mass supply rates from the envelope and angular momentum of the gas infalling onto the disk. Moreover, we have investigated to what extent the non-steady model explains the evolution observed in a high-resolution 3D simulation, where the disk fragmentation occurs when the protostellar mass is less than $1~M_\odot$.
\par
We have first investigated the fiducial case, where we assume the typical envelope structure suggested by 1D hydrodynamical calculations following the cloud collapse \citep{Omukai_and_Nishi_1998}. Our model shows that a self-gravitating disk grows and evolves around the star. The gravitational torque regulates the mass and angular momentum transfer, so that the Toomre Q parameter is close to unity almost everywhere in the disk. We have shown that the accretion rate onto the star is an order of magnitude smaller than the mass supply rate from the envelope.
As a result, the disk becomes more massive than the star before the star accretes the gas of $0.1~M_\odot$. Such a feature has not been seen in the previous steady disk models \citep[e.g.,][]{Tanaka_and_Omukai_2014}.
\par
To understand why only a small fraction of the gas supplied from the envelope is transferred to the star through the disk, we have also developed an analytical model. We show that a large part of the mass supply from the envelope is used to extend the outer edge of the disk.
Such evolution occurs because the gas infalling onto the disk later has higher angular momentum, which is predestined by the structure of the envelope. The analytical argument predicts that the accretion rate onto the star is $\sim 10$\% of the mass supply rate from the envelope, which agrees well with our 1D model results.
\par
We have next studied the parameter dependence of the evolution of the disk in the 1D model, varying the mass supply rate from the envelope and its rotation. The results show that the disk is more gravitationally unstable with the higher mass supply rate, as already shown in previous studies \citep{Matsukoba_et_al_2019}.
The disk structure is almost independent of the rotational degree of the envelope, except that the disk size is larger with the higher angular momentum of the envelope. Despite some variations, all these cases show that the the accretion rate onto the star is much lower than the mass supply rate from the envelope. The disk-to-star mass ratio $\xi$ generally exceeds the unity before the star accretes the gas of $\sim1~M_\odot$.
\par
To examine the validity of our 1D model, We have further compared it to a 3D hydrodynamics simulation. To this end, we have newly performed a high-resolution simulation that follows the early evolution in the protostellar accretion stage using a modified version of the adaptive mesh refinement code SFUMATO-RT \citep{Matsumoto_2007,Sugimura_et_al_2020}.
Our simulation run shows that the disk fragmentation occurs before the stellar mass exceeds $1~M_\odot$, as reported in the previous simulations \citep[e.g.,][]{Clark_et_al_2011}.
The disk-to-star mass ratio $\xi$ reaches $\simeq 4$ just before the fragmentation, which well agrees with the evolution shown by the 1D model. The overall structure of the self-gravitating disk also matches between the simulation and model, though the temperature in the simulation is higher than that in the model by a factor of $\simeq 2$.
\par
We have shown that in both the 1D non-steady model and 3D simulation the massive disk commonly appears soon after the birth of the protostar.
We argue that the formation of such a massive disk leads to the early disk fragmentation, as generally seen in the previous 3D simulations.
Intriguingly, the formation of the disk more massive than the central star is also shown to occur in the present-day star formation \citep[e.g.,][]{Inutsuka_et_al_2010}, regardless of the significant difference in the thermal evolution of the star-forming gas. Comprehensive understanding of the massive disk formation, including the primordial and present-day cases, is a task for further studies.
\acknowledgments
The authors thank Sanemichi Takahashi, Kenji Eric Sadanari, Kazuhiro Shima, Ryoki Matsukoba, Kei Tanaka, Kazuyuki Omukai, Kunihito Ioka, and Takahiro Tanaka for fruitful discussions and comments.
The authors also appreciate Tomoaki Matsumoto for great contribution to the code development.
KS appreciates the support by the Fellowship of the Japan Society for the Promotion of Science for Research Abroad.
The numerical simulations were performed on the Cray XC50 (Aterui II) at the Center for Computational Astrophysics (CfCA) of National Astronomical Observatory of Japan.
This work is financially supported by the Grants-in-Aid for Basic Research by the Ministry of Education, Science and Culture of Japan (17H06360, 19H01934: T.H.).

\bibliographystyle{aasjournal}
\appendix
\section{Derivation of Equation \plural} \label{app:analytic}
To derive Equation \plural, we rewrite the second term in the right-hand side in Equation \eqref{eq:Mdot_ratio}. To this end, we transform the quantities contained in the second term, i.e., $\dot{M}_{\mr{e,tot}}$, $c_s$, $\Omega$, and $d\varpi_{\mr{d}}/dt$, considering the structure of the accretion envelope and marginally stable disk.
\par
We suppose that the enclosed gas inside the radius $r_{\mr{in}}$ of the original accretion envelope has accreted onto the disk by the epoch considered.
Assuming the mass distribution in the envelope given by Equation \eqref{eq:Larson_rho} with $\gamma = 1.1$, we describe the total mass of the accreted gas $M_{r_{\mr{in}}}$ as
\begin{eqnarray}
 M_{r_{\mr{in}}} &=& M_{*,\mr{ini}} + \int_{r_*}^{r_{\mr{in}}} 4\pi r_{\mr{in}}^2 \rho dr \nonumber \\
 &\simeq& \f{36\pi}{7} C_1 K^{10/9} r_{\mr{in}}^{7/9} \h , \label{eq:M_rin}
\end{eqnarray}
where $M_{*,\mr{ini}}$ and $r_*$ are the initial protostellar mass and radius, and for the second equality we assume that $M_{r_{\mr{in}}} \gg M_{*,\mr{ini}}$ and $r_\mr{in} \gg r_*$.
We define the radius of the outer edge of the disk as $\varpi_{\mr d}$, similarly as in Section \ref{sec:mass_ratio}. The enclosed mass $M_{\varpi_{\mr d}}$ is thus given by
\begin{eqnarray}
 M_{\varpi_{\mr{d}}} = M_{r_{\mr{in}}} \label{eq:M_varpi_rin} \h .
\end{eqnarray}
From the angular momentum conservation during the infall from the envelope onto the disk,
\begin{equation}
   \sqrt{\varpi_\mr{d} G M_{\varpi_\mr{d}}} = f_\mr{Kep} \sqrt{r_\mr{in} G M_{r_\mr{in}}} , \nonumber
\end{equation}
$\varpi_{\mr{d}}$ and $r_\mr{in}$ are related as follows:
\begin{eqnarray}
 \varpi_\mr{d} = f_\mr{Kep}^2 r_\mr{in} \h . \label{eq:varpi_rin}
\end{eqnarray}
With Equations \eqref{eq:disk_Omega}, \eqref{eq:M_rin}, \eqref{eq:M_varpi_rin}, and \eqref{eq:varpi_rin}, the angular velocity $\Omega$ at the disk outer edge is written as
\begin{eqnarray}
 \Omega = 6 \sqrt{\f{\pi G}{7}} C_1^{1/2} K^{5/9} f_\mr{Kep}^{-7/9} \varpi_\mr{d}^{-10/9} \h. \label{eq:Omega_analytic}
\end{eqnarray}
We can get the relation of $t$ and $r_\mr{in}$ from Equations \eqref{eq:Larson_v} and \eqref{eq:t_infall}, and $dr_\mr{in}/dt$ is given by
\begin{eqnarray}
 \f{dr_\mr{in}}{dt} = \f{9}{10} C_2 K^{5/9} r_\mr{in}^{-1/9} \h . \label{eq:dr_dt}
\end{eqnarray}
From Equation \eqref{eq:varpi_rin}, we have
\begin{eqnarray}
 \f{d\varpi_\mr{d}}{dt} = \f{9}{10} C_2 K^{5/9} f_\mr{Kep}^{20/9} \varpi_{\mr d}^{-1/9} . \label{eq:dvarpi_dt}
\end{eqnarray}
With Equations \eqref{eq:Larson_rho}, \eqref{eq:Mdot_e_tot}, \eqref{eq:varpi_rin}, and \eqref{eq:dr_dt}, the total mass supply rate from the envelope $\dot{M}_{\mr{e,tot}}$ is given by
\begin{eqnarray}
 \dot{M}_{\mr{e,tot}} = \f{18\pi}{5} C_1 C_2 K^{5/3} f_\mr{Kep}^{2/3} \varpi_\mr{d}^{-1/3} \h . \label{eq:Mdot_e_tot_analytic}
\end{eqnarray}
In the marginally stable disk with $Q\sim Q_{\mr{crit}}$, from Equation \eqref{eq:Q_def} the surface density is
\begin{eqnarray}
 \Sigma \sim \f{\Omega c_s}{\pi G Q_\mr{crit}} \h.
\end{eqnarray}
With Equations \eqref{eq:nH_rho_relation} and \eqref{eq:iso-Sigmarho}, the number density of hydrogen nuclei at the disk edge is thus given by
\begin{eqnarray}
 n_{\chH} &=& \f{1}{m_{\chH}}\f{\Sigma}{\sqrt{2\pi}H} \nonumber \\
 &\sim& \f{M_{\varpi_\mr{d}}}{\sqrt{2\pi^3}m_{\chH}Q_\mr{crit}\varpi_\mr{d}^3} \h ,
 \label{eq:nH_edge}
\end{eqnarray}
where we neglect $y_\mr{He}$ in Equation \eqref{eq:nH_rho_relation}.
The sound velocity $c_s$ is written as $c_s = c_{s,1000} (T/1000~\mr{K})^{1/2}$ where $c_{s,1000}=3.3\times10^5~{\rm cm/s}$ is the sound velocity at $T = 1000$~K.
With Equations \eqref{eq:T_edge} and \eqref{eq:nH_edge}, we can rewrite $c_s$ as
\begin{eqnarray}
 c_s  = 3.7 \times 10^5 \bra{\f{Q_\mr{crit}}{1.0}}^{-1/20} \bra{\f{f_T}{0.4}}^{1/2} \bra{\f{K}{K_\mr{fid}}}^{5/9}  \bra{\f{f_\mr{Kep}}{0.5}}^{-7/90} \bra{\f{\varpi_\mr{d}}{1\mr{AU}}}^{-1/9} \hspace{2mm} \mr{cm/s} \h , \label{eq:c_s_disk}
\end{eqnarray}
Substituting Equations \eqref{eq:Omega_analytic}, \eqref{eq:dvarpi_dt}, \eqref{eq:Mdot_e_tot_analytic}, and \eqref{eq:c_s_disk} into Equation \eqref{eq:Mdot_ratio}, we finally obtain
\begin{eqnarray}
 \f{\dot{M}_{\mr{d,edge}}}{\dot{M}_{\mr{e,tot}}}
 = 1 &-& 0.68 \bra{\f{Q_{\mr{crit}}}{1.0}}^{-21/20} \bra{\f{f_T}{0.4}}^{1/2}  \bra{\f{f_{\mr{Kep}}}{0.5}}^{7/10} \h .
\end{eqnarray}
\end{document}